\documentclass[useAMS,usenatbib,a4]{mn2e}

\usepackage{amssymb}
\usepackage[fleqn]{amsmath}
\usepackage{times}
\usepackage{graphicx}
\usepackage{multirow}

\title[COSMOS: EC dominated X-ray sources]
{Extended X-ray emission from non-thermal sources in the COSMOS field: A detailed study of a large radio galaxy at $z=1.168$}

\author[Jeli{\'c} et al. ]{Vibor Jeli{\'c}$^{1}$\thanks{E-mail: 
jelic@astron.nl}, Vernesa Smol{\v c}i{\'c}$^{2,3,4,5}$, Alexis Finoguenov$^{6,7}$, Masayuki Tanaka$^{8}$,
\newauthor Francesca Civano$^{9}$, Eva Schinnerer$^{10}$, Nico Cappelluti$^{11}$, and Anton Koekemoer$^{12}$\\
$^{1}$ASTRON, the Netherlands Institute for Radio Astronomy, P.O. Box 2, 7990 AA Dwingeloo, the Netherlands\\ 
$^{2}$ESO ALMA COFUND Fellow\\
$^{3}$Argelander Institut for Astronomy, Auf dem H$\ddot{u}$gel 71, Bonn, 53121, Germany\\
$^{4}$European Southern Observatory, Karl-Schwarzschild-Strasse 2, 85748 Garching b. M$\ddot{u}$nchen, Germany\\
$^{5}$University of Zagreb, Physics Department, Bijeni\v{c}ka cesta 32, 10002 Zagreb, Croatia\\
$^{6}$Max-Planck-Institut f$\ddot{u}$r extraterrestrische Physik, Giessenbachstrasse, 85748 Garching b. M$\ddot{u}$nchen, Germany\\
$^{7}$University of Maryland, Baltimore County, 1000 Hilltop Circle, Baltimore, MD 21250, USA\\
$^{8}$Institute for the Physics and Mathematics of the Universe, The University of Tokyo, 5-1-5 Kashiwanoha, Kashiwa-shi, Chiba 277-8583, Japan\\
$^{9}$Harvard-Smithsonian Centre for Astrophysics, 60 Garden Street, Cambridge, MA 02138, USA\\
$^{10}$Max-Planck-Institut f$\ddot{u}$r Astronomie, K$\ddot{o}$nigstuhl 17, D-69117 Heidelberg, Germany\\
$^{11}$INAF-Osservatorio Astronomico di Bologna, Via Ranzani 1, I - 40127 Bologna, Italy\\
$^{12}$Space Telescope Science Institute, 3700 San Martin Drive, Baltimore MD 21218, U.S.A.}

\begin{document}


\pagerange{\pageref{firstpage}--\pageref{lastpage}} \pubyear{2012}

\maketitle

\label{firstpage}
\begin{abstract}
X-ray selected galaxy group samples are usually generated by searching for extended X-ray sources that reflect the thermal radiation of the intragroup medium. 
On the other hand, large radio galaxies that regularly occupy galaxy groups also emit in the X-ray window, and their contribution to X-ray selected group
 samples is still not well understood. In order to investigate their relative importance, we have carried out a systematic search for non-thermal extended 
 X-ray sources in the COSMOS field. Based on the morphological coincidence of X-ray and radio extensions, out of 60 radio galaxies, and $\sim300$
  extended X-ray sources, we find only one candidate where the observed extended X-ray emission arises from non-thermal processes related to radio 
  galaxies. We present a detailed analysis of this source, and its environment. Our results yield that external Inverse Compton emission of the lobes 
  is the dominant process that generates the observed X-ray emission of our extended X-ray candidate, with a minor contribution from the gas 
  of the galaxy group hosting the radio galaxy. Finally, we show that finding only one potential candidate in the COSMOS field (in a redshift range 
  $0<z<6$ and with radio luminosity between $10^{25}$ and $10^{30}$ ${\rm W/Hz}$) is consistent with expected X-ray--counts arising from synchrotron lobes. 
  This implies that these sources are not a prominent source of contamination in samples of X-ray selected clusters/groups, but they could potentially dominate the
   $z>1$ cluster counts at the bright end ($S_{\rm X}>7\cdot10^{-15}~{\rm erg~s^{-1}~cm^2}$).
\end{abstract}


\section{Introduction}\label{sec:intro}

Radio galaxies constitute the most powerful and the largest-scale active galactic nuclei phenomena. Their jets span distances of up to 1~Mpc. The optical hosts of powerful radio loud AGN usually have colors consistent with that of galaxies in the so called ``green valley'', i.e., a region in a color vs.\ stellar mass plane thought to reflect the evolutionary transition of blue-star forming galaxies to red-and-dead ones \citep{smolcic09b}.  They have a higher molecular gas content relative to less powerful radio galaxies inhabiting the red sequence \citep{smolcic11}. They often show signs of merger activity \citep{simpson02}, and they usually reside in group environments \citep{baum92}. As their high brightness in the radio allows detections out to high redshift, powerful radio galaxies are often used as tracers of distant groups/clusters of galaxies. 

A direct, and powerful way to detect clusters and groups of galaxies is via extended X-ray emission reflecting thermal bremstrahlung radiation from the cluster/group gas (i.e.\ intercluster/intragroup medium; ICM). With the advent of deep X-ray surveys, and the optimization of cluster/group-detection algorithms hundreds of faint X-ray clusters have been detected in such a way out to $z\sim2$ \citep{finoguenov07, finoguenov09, finoguenov10, bielby10, tanaka10,henry10,gobat11}. However, large radio galaxies, which are preferentially found in such galaxy groups/clusters, are also luminous in the X-rays, thus possibly biasing the X-ray cluster searches (see below). Furthermore, the X-ray emission of their jets, hot-spots and lobes may outlive the radio emission, as IC-CMB emission can downgrade the electrons required for (high frequency) radio emission \citep{fabian09,mocz11}.  

It has been demonstrated that the X-ray emission of radio galaxies arises from three main processes: synchrotron radiation emitted by relativistic electrons, inverse Compton (IC) scattering of the synchrotron photons (so called synchrotron self-Compton, SSC, process) or the cosmic microwave background (CMB) photons (so called external Compton, EC, process) off the relativistic electrons. The first and second processes are predominant in jet-knots and hot-spots of low-redshift radio galaxies, while the last process can regularly be associated with the extended lobes of radio galaxies (see \citealt{kataoka05} for details).  It is important to note that the X-ray flux of EC emission is proportional to the energy density of the CMB photons that rises with redshift as $(1+z)^4$. This rise in flux partially compensates the dimming effect due to the large distances to such sources \citep{felten69, schwartz02}, thereby making them detectable also at high redshifts \citep[$z\gtrsim1$, e.g.,][]{simpson02, fabian03, blundell06, johnson07, erlund08}.  

Depending on the expected number of extended X-ray--emitting radio lobe sources (via EC), they might be a serious contaminant in deep X-ray surveys searching for clusters/groups at high redshifts. Assuming that radio AGN emit in the X-ray via EC emission, and evolving radio AGN luminosity functions \citet{celotti04} argued that at redshifts of $z\gtrsim1$ the EC emission of radio lobes will likely dominate in X-rays (at $L_X>10^{44}$~erg/s) over the thermal gas emission of the clusters/groups. 
When, however, a radio galaxy lies in a cluster/group, the total X-ray emission from this system is likely a combination of EC emission from the radio lobes of the galaxy, and thermal bremsstrahlung emission from the cluster/group gas. However, so far there has been just a single survey \citep{finoguenov10} that tried to quantify that.  
\citet{finoguenov10} searched for X-ray galaxy clusters in the Subaru-XMM Deep Field (SXDF), covering $1.3\Box^\circ$. They identified 57 cluster candidates, 4 of which they consider likely EC X-ray sources and 2 more which are likely a combination of EC scattering and thermal cluster emission.

In order to put stronger constraints on extended X-ray sources due to non-thermal processes in deep X-ray surveys, here we present a search for EC dominated X-ray candidates in the Cosmic Evolution Survey \citep[COSMOS,][]{scoville07} field, and discuss their expected abundance in future X-ray surveys. The paper is organized as follows. In Sec.~\ref{sec:data} we
give a brief overview of the COSMOS survey and the data used for this
work.  In Sec.~\ref{sec:search} we describe our systematic search for EC dominated X-ray sources
in the COSMOS field. The radio, X-ray, and optical properties of our single candidate are presented 
in Sec.~\ref{sec:Xraylobe}, while the properties of the associated galaxy group are discussed in Sec.~\ref{sec:group}. 
The origin of observed X-ray emission is discussed in Sec.~\ref{sec:Xrayexp}. Section~\ref{sec:expcounts} presents predictions for
EC dominated X-ray sources in future X-ray surveys. We conclude the paper with Sec.~\ref{sec:con}.

Throughout we assume $\Lambda$CDM-cosmology with WMAP7 parameters
\citep{komatsu11}: $h=0.71$, $\Omega_{\rm b}=0.045$,
$\Omega_{\rm m}=0.27$ and $\Omega_\Lambda=0.73$, and we use AB magnitude system.

\section{Data}\label{sec:data}
The COSMOS survey is designed to probe the formation and evolution of galaxies as a function of cosmic time and large scale structure environment. 
The survey covers a $2\Box^\circ$ area close to celestial equator \citep{scoville07} with multi-wavelength imaging from X-ray to radio wavelengths, 
including HST/ACS imaging \citep[see][]{koekemoer07} and optical spectroscopy \citep[$z$COSMOS,][]{lilly07,lilly09}. 
Due to this broad wavelength coverage the photometric redshifts for galaxies in the COSMOS field are determined to 
an excellent accuracy of $0.007 \cdot (1+z)$ for $i^+<22.5$ \citep{ilbert09, salvato09}. The stellar masses of every source
 in the photometric redshift catalog are estimated using a Salpeter initial mass function (IMF). 

In radio wavelengths, the COSMOS field has been observed at $1.4~{\rm GHz}$
\citep[VLA-COSMOS survey,][]{schinnerer07, schinnerer10}  and $327~{\rm
  MHz}$ (Smol{\v c}i{\'c} et al., {\em in prep.})  with the NRAO
Very Large Array (VLA) in A and C configurations. The reached {\em
  rms} and resolution at $327~{\rm MHz}$ ($1.4~{\rm GHz}$) are $\sim
0.4~{\rm mJy~beam^{-1}}$ ($\sim 8~{\rm \mu Jy~beam^{-1}}$) and
$6.0''\times 5.4''$ ($1.5''\times1.4''$), respectively. The 1.4~GHz
catalog utilized here contains $\sim2400$ ($\geq5\sigma$) sources, $60$ of which are radio
galaxies (with clear core/jet/lobe features).

X-ray observations of the COSMOS field have been performed both with
{\em XMM-Newton} \citep[1.5 Ms covering $2\Box^\circ$,][]{hasinger07}
and {\em Chandra} \citep[1.8 Ms covering inner
$1\Box^\circ$,][]{elvis09}.  Based on a composite mosaic of both
observations, a galaxy group catalogue has been generated
\citep[][Finoguenov et al., in prep.]{finoguenov07}. It contains $\sim300$ extended sources out to redshifts of 1.3 with total masses within a radius at 200 times the critical density in the range of $M_{200}\in[7\times10^{12}, 3\times10^{14}]~\mathrm{M_\odot}$.  The extended X-ray source detection is based on a wavelet analysis technique and includes removal of point sources \citep{finoguenov09}. Each X-ray cluster candidate has been further independently verified via an optical galaxy cluster search (making use of both the COSMOS photometric and spectroscopic redshifts, following the procedure outlined in \citealt{finoguenov10}).

\section{Search for joint radio/X-ray coexistance in the COSMOS field}\label{sec:search}
Using the data mentioned in Sec.\ref{sec:data}, we carried out a systematic search for EC dominated X-ray sources. 
The search is based on the coincidence of radio and X-ray emissions. We compare the centering and positional angle of each of the 60 morphologically 
complex radio sources drawn from the 20~cm VLA-COSMOS catalog with the associated extended X-ray emission in the $0.5$--$2~{\rm keV}$ band. To avoid a possible identification  
bias towards availability of an optical counterpart no prior on either source has been applied.

\begin{figure*}
\centering \includegraphics[width=.45\textwidth]{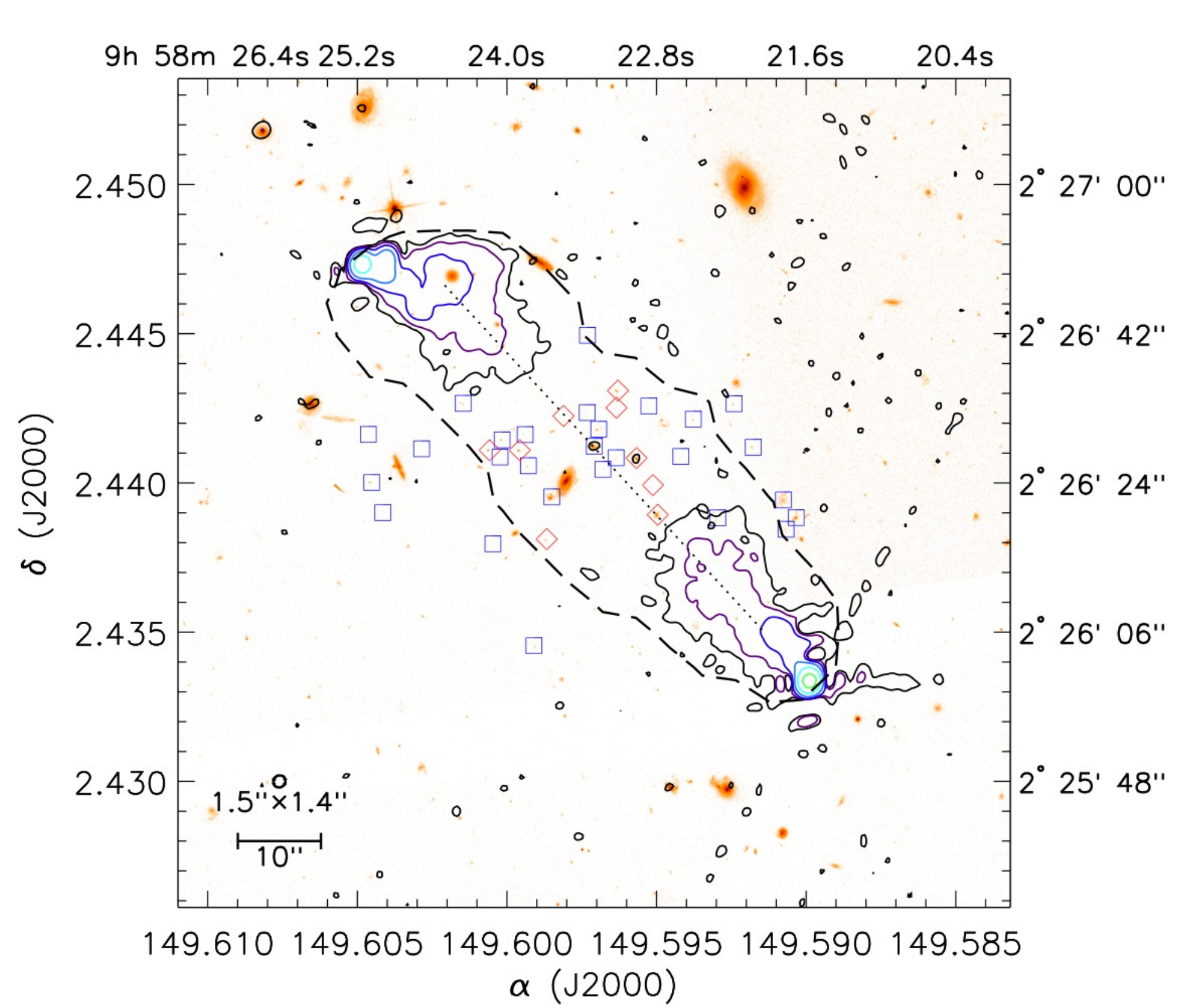}
\centering \includegraphics[width=.45\textwidth]{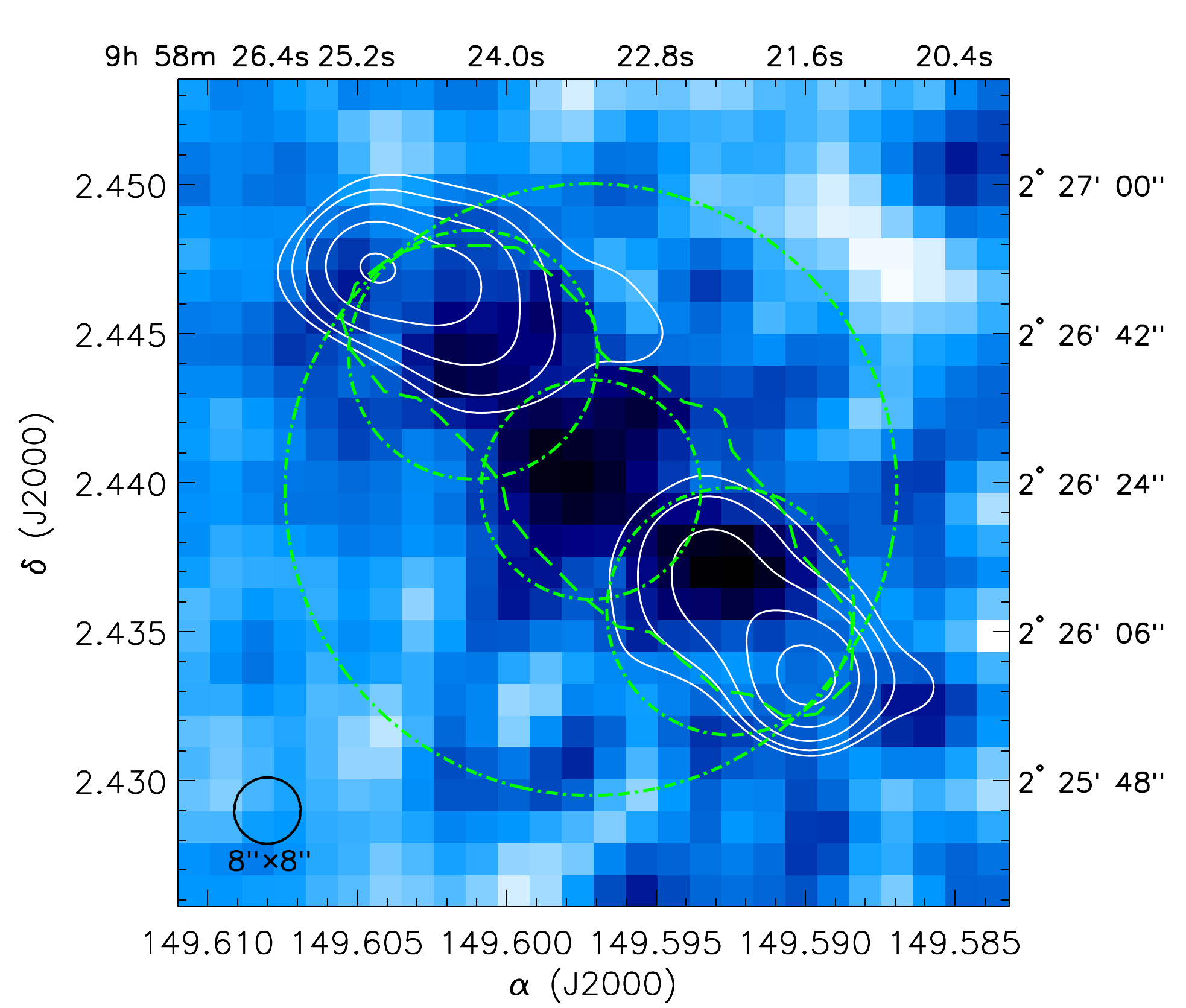}
\caption{Multi-wavelength image of the region around the EC dominated X-ray candidate. \emph{Left:}  HST-ACS {\emph i}-band image is presented in a red scale. 
The radio emission of the associated radio galaxy at $1.4~{\rm GHz}$ is plotted with solid $3^i\sigma$ level contours (where $i=1,2,...$ and $\sigma=17.16~{\rm \mu Jy}/beam$).
The synthesized beam of the radio image is $1.5''\times1.4''$ (thick solid ellipse plotted in the bottom left corner). Borders of extended X-ray emission obtained 
by the wavelet analysis of the XMM-\emph{Newton} image are shown by the dashed line.  Associated optical/IR galaxies identified via the Voronoi method 
are marked with blue boxes and red diamonds, according to their $i-K_s$ color (see Sec.~\ref{sec:group}). \emph{Right:} XMM-\emph{Newton}, 
background point-source subtracted and smoothed to $8''\times8''$ resolution, image of the region around the EC dominated X-ray candidate 
(presented in a blue scale; note that darker colors show regions with stronger X-ray emission). The radio emission at $1.4~{\rm GHz}$ is smoothed to the resolution of the X-ray image and 
overplotted as solid white contours. A dashed green line marks the X-ray flux extraction region associated with the extended X-ray emission, while the dash-dotted green circles 
are regions associated with the core, lobes, and a putative group emission.}
\label{fig:Xlobe}
\end{figure*}

In addition to a positional match, a coincidence in elongation of  radio and X-ray emissions (to within $10^\circ$) is required by the method. This creates a robust identification of the extended X-ray source as a counter-part of radio-lobes, as the chance probability of such an alignment is $10^{-4}$, given the density of both X-ray and radio sources, and chance alignment of principal axes. However, such a search is limited to the early phase of radio-activity and may not select X-ray only lobes.  
Compared to a similar study done on the $1.3\Box^\circ$ of  SXDF survey \citep{finoguenov10}, we find just one EC dominated X-ray candidate (Fig.~\ref{fig:Xlobe}), which implies a factor of 6 lower frequency of the phenomenon as inferred by \citet{finoguenov10}. We note that the redshift of the COSMOS source, $z_{\rm spec}=1.168$, is at a similar redshift as the brightest EC dominated X-ray candidate in SXDF.

Hereafter, we refer to this system as the EC dominated X-ray candidate. Its radio, X-ray, and optical properties are presented in the following section.

\section{EC dominated X-ray candidate}\label{sec:Xraylobe}

\subsection{X-ray properties}\label{sec:Xray}
The extended X-ray emission associated with our candidate has been identified using the full XMM-\emph{Newton} survey of the COSMOS field 
(it is outside the area covered by \emph{Chandra}).
However, the best imaging quality is available in the observation 0203362201, which has a $26~{\rm ks}$ cleaned exposure and the object is located next to the telescope optical axis
(best PSF and sensitivity). We show the X-ray counts, smoothed to $8''\times8''$ resolution, in Fig.~\ref{fig:Xlobe}. 
From the figure it is obvious that the X-ray emission is arising from the region associated with the radio galaxy. 
The emission shows both unresolved and resolved components with a total flux of $(3.6\pm1.2)\cdot 10^{-15}~{\rm erg~s^{-1}~cm^{-2}}$ in the observed 0.5-2 keV band  
(dashed line in left panel of Fig.~\ref{fig:Xlobe}).

\begin{figure}
\centering \includegraphics[width=.45\textwidth]{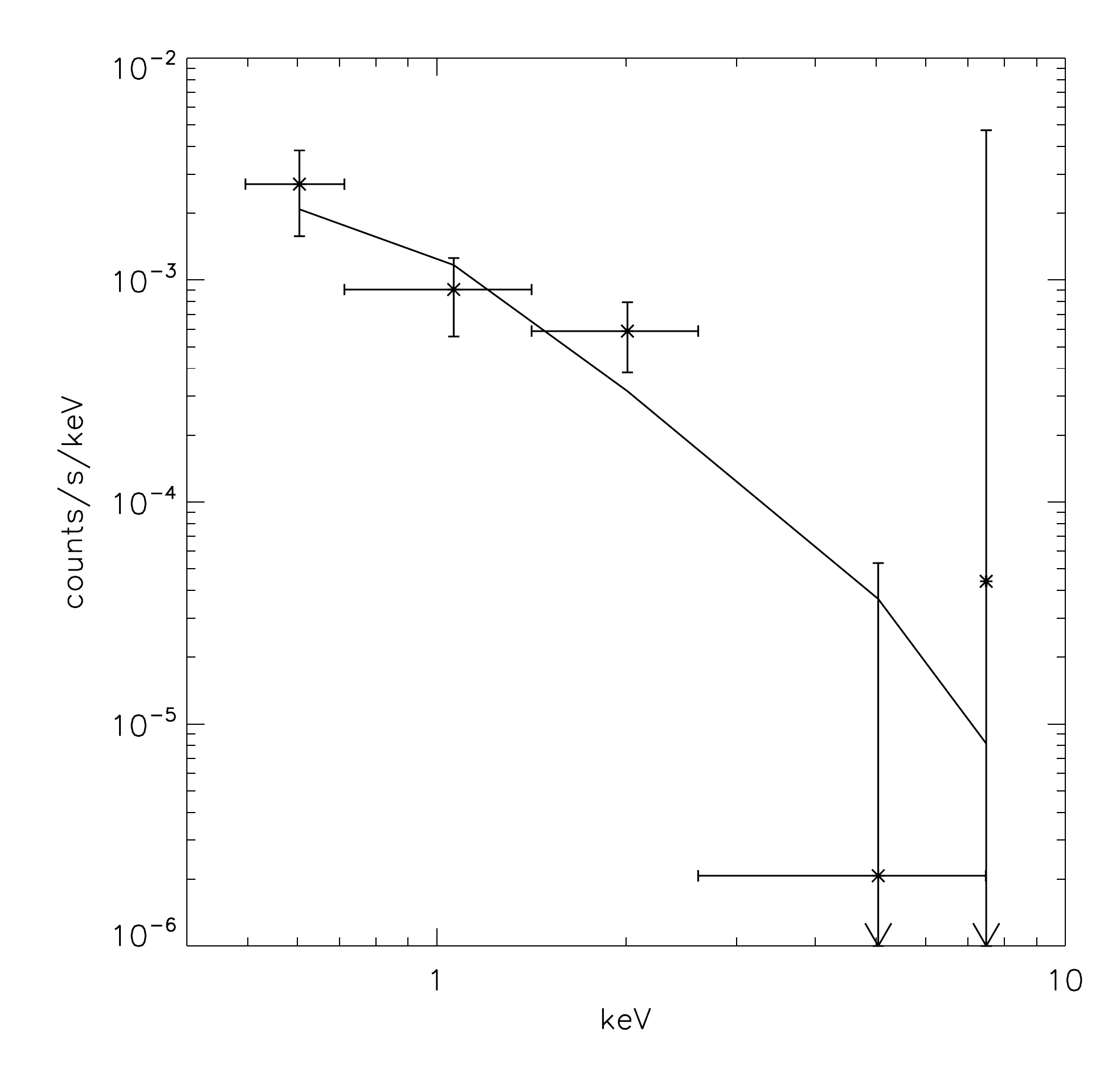}
\caption{Fitted power-law spectrum of the X-ray emission in the $0.4-10~{\rm keV}$ range using both the stowed background and
the local background.  Data points are given for the X-ray emission associated with the lobes (see Fig.~\ref{fig:Xlobe}). 
Note that the spectral shape is hard, which is consistent with the non-thermal emission from
EC scattering, rather than thermal emission arising from the galaxy group.}
\label{fig:specfit}
\end{figure}

In order to get a better insight into the X-ray properties associated with the source, we have extracted the counts from three zones within the X-ray map. These zones, illustrated by the dash-dotted green areas in the right panel of Fig.~\ref{fig:Xlobe}, encompass the core, both lobes, and a putative group emission. 
Using both the stowed background and the local background region, selected from the same chip, but 2 arcmin away from the source, we analyzed the spectra
in the $0.4-7.5~{\rm keV}$ range for each zone separately. Significant detections have been obtained in all three zones. Using the XMM PSF model, and for the putative group emission an  assumption of  azimutal symmetry, we have solved for intrinsic flux of all these components: in the observed 0.5-2 keV band and in units of $10^{-15}$ erg s$^{-1}$ cm$^{-2}$ the lobe emission amounts to  $2.7\pm0.7$, core emission to $1.7\pm0.4$, total group emission of $1.4\pm1.2$ and its contribution to the lobes is $0.5\pm0.4$. The group component has not been significantly detected. 
In Fig.~\ref{fig:specfit} we show the spectrum of the lobes. Given the marginal contribution of the group emission, established above, we ignored it,  finding a photon index of
 $2.6\pm0.5$ ($\chi^2=3.2/7~{\rm dof}$). The core shows a hard spectrum with a photon index of $0.6\pm0.4$.

\subsection{Radio properties}\label{sec:radio}
The radio counterpart of the EC dominated X-ray candidate is identified as a large powerful radio galaxy J095822.93+022619.8 with integrated source flux density of $S_{1.4~{\rm GHz}}=(112.9\pm0.5)~{\rm mJy}$ \citep{schinnerer07} and $S_{327~{\rm MHz}}=(470.6\pm0.5)~{\rm mJy}$ (Smol{\v c}i{\'c} et al., {\em in prep.}). The spectroscopic redshift of its optical counterpart is $1.1684\pm0.0007$ (see Sec.~\ref{sec:opti}), which corresponds to a luminosity distance of $d_L=8062~{\rm Mpc}$ for the adopted cosmology. Thus, the monochromatic radio powers
of the source at 1.4~GHz and 327~MHz are $L_{1.4~{\rm GHz}}=8.8\cdot10^{26}~{\rm W/Hz}$ and $L_{327~{\rm MHz}}=3.7\cdot10^{27}~{\rm W/Hz}$, respectively.

Morphologically, the source shows a structure of Fanaroff-Riley (FR) class II \citep{fanaroff74}, i.e., a luminous radio galaxies with predominant emission from the lobes (see Fig.~\ref{fig:Xlobe} and Table~\ref{tab:radiomorph}). The radio emission of the jets themselves is not detected. The radio core, located centrally between the lobes (see the line joining the lobes in Fig.~\ref{fig:Xlobe}) is detected at $\sim4\sigma$ in the 1.4~GHz radio map. The central parts of the lobes are  $\sim26.5~{\rm arcsec}$ ($\sim220~{\rm kpc}$) away from the core in two diametrically opposite directions (NE-SW direction on the plane of the sky). The hotspot of the NE lobe is off the joint line between the lobes, suggesting interaction with a medium.

\begin{table}
  \centering
  \caption{Integrated flux density and projected angular size of morphological features resolved in the 1.4~GHz radio map.}
  \begin{tabular}{lcc}\hline\hline
{\sc morphological features} & $S_{\rm 1.4~GHz}~[{\rm  mJy}]$ & {\sc angular size}   \\  \hline
{\sc NE hotspot} & $19.5\pm0.5$ & $1.5''\times1.4''$ \\
{\sc SW hotspot} & $35.3\pm0.5$ & $1.5''\times1.4''$ \\
{\sc NE lobe} & $36.2\pm0.5$ & $23.5''\times16.4''$ \\
{\sc SW lobe} & $21.9\pm0.5$ & $24.1''\times14.3''$ \\ \hline
\hline
  \end{tabular}\label{tab:radiomorph}
\end{table}

\begin{figure*}
\centering \includegraphics[width=.33\textwidth]{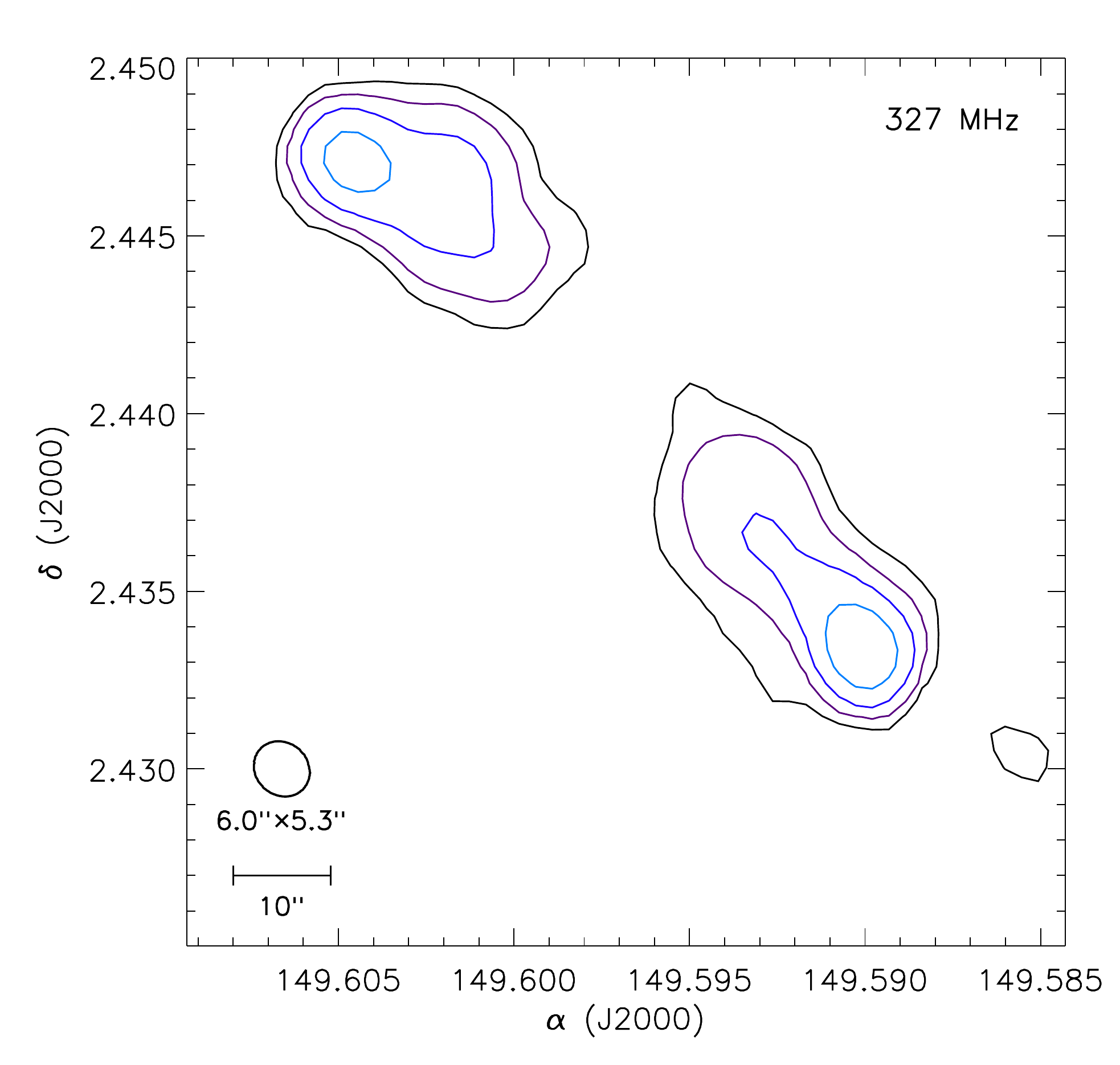}
\centering \includegraphics[width=.33\textwidth]{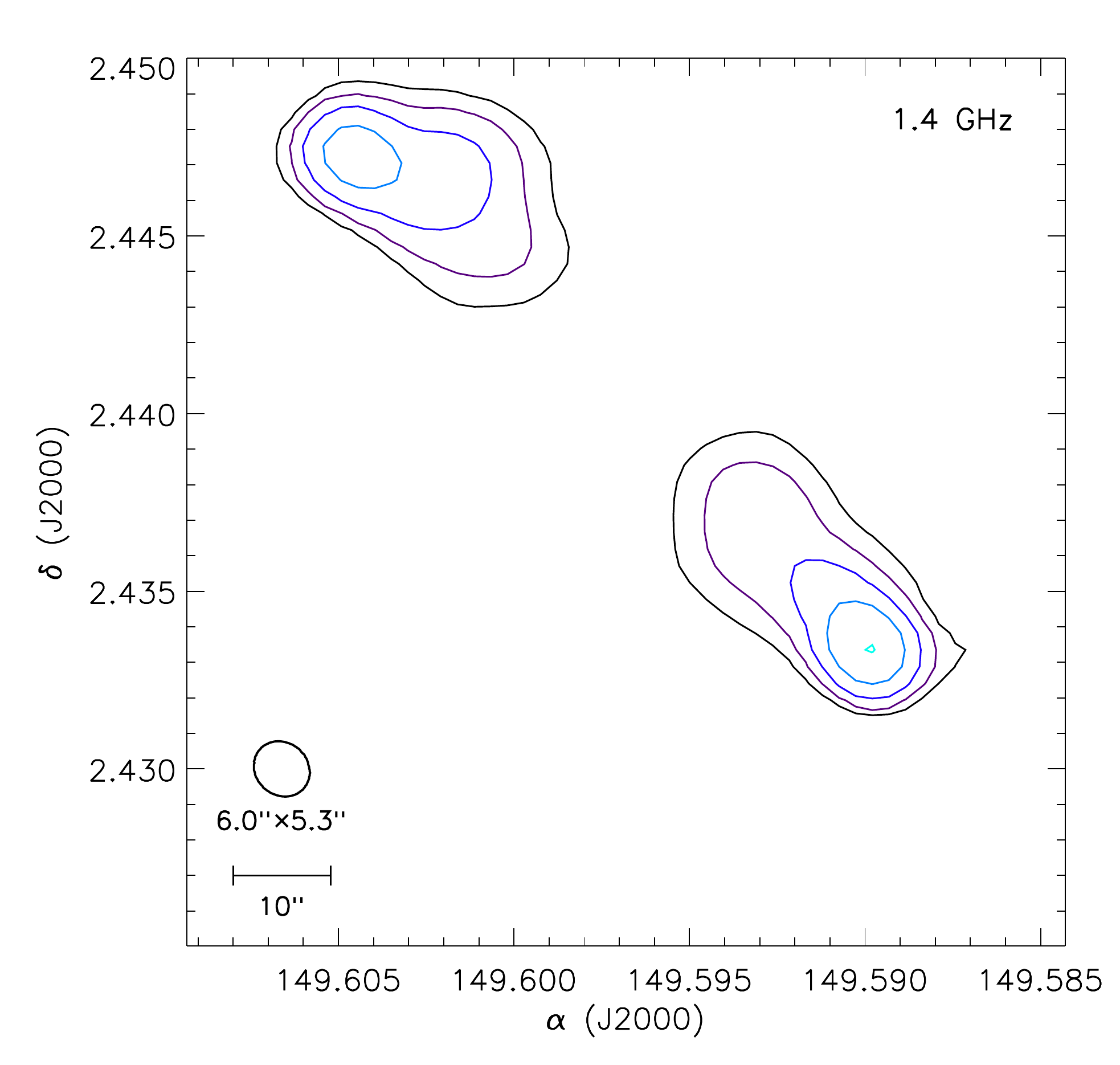}
\centering \includegraphics[width=.33\textwidth]{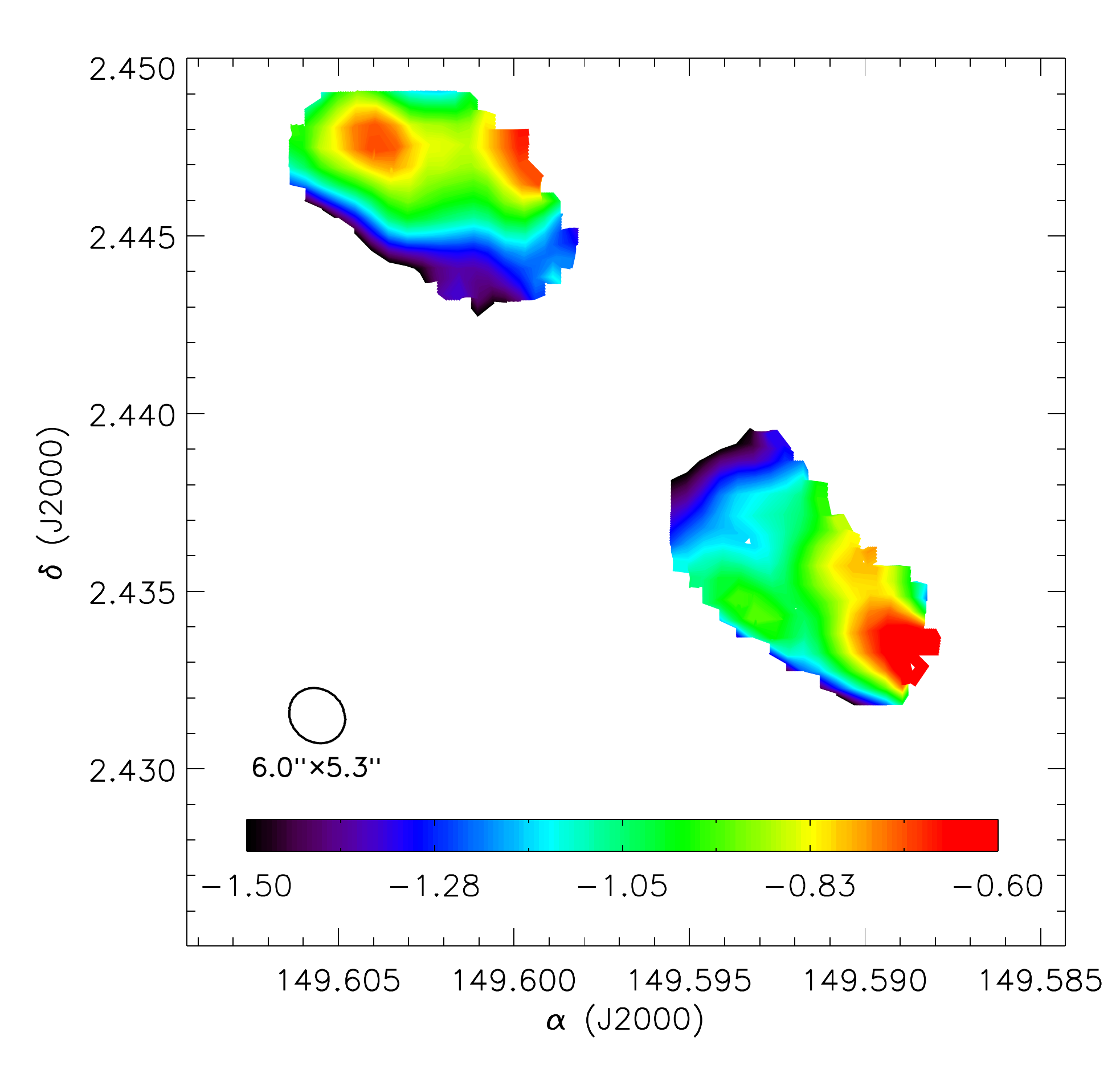}
\caption{The radio emission of the associated radio galaxy at $327~{\rm MHz}$ (\emph{left}), at $1.4~{\rm GHz}$  (\emph{middle}) convolved to the resolution of the $327~{\rm MHz}$ image, 
and the corresponding spectral index map of the radio emission ($S_\nu \propto \nu^{\alpha_r}$; \emph{right}). Contours of the radio emission
have $3^i\sigma$ levels, where $i=1,2,...$, $\sigma_{327~{\rm MHz}}=0.4~{\rm  mJy}/beam$, and $\sigma_{1.4~{\rm GHz}}=0.14~{\rm  mJy}/beam$. 
The synthesized beams of the radio images are $6.0''\times5.4''$ (thick solid ellipses plotted in the bottom left corner of the images). 
As expected, the spectral index steepen radially towards the core.  The average spectral index of both lobes together is $\alpha_r=-(1.0\pm0.2)$. 
By comparing the spectral index map with the X-ray emission shown in Fig.~\ref{fig:Xlobe}, we note that the X-ray emission stems from the zones of steeper spectral indexes.}
\label{fig:specind}
\end{figure*}

Combining the images at $327~{\rm MHz}$ and  $1.4~{\rm GHz}$, the spectral index map of the source is obtained (see Fig.~\ref{fig:specind}; $S_\nu \propto \nu^{\alpha_r}$). Prior to the spectral index calculation (i) the $1.4~{\rm GHz}$ image was convolved and re-gridded to match the resolution and  pixel scale of the $327~{\rm MHz}$ image, (ii) positional alignment  between the two images was checked using 45 point sources (Smol{\v c}i{\'c} et al., {\em in prep.}); and  (iii) pixels with values below $3\sigma$ in each image were blanked. The spectral index shows the expected behavior for both lobes. It steepens radially towards the core from $-0.6$ to $-1.5$. The average spectral index is $\alpha_r=-(1.0\pm0.2)$. For clarity, we also calculate the spectral index from the integrated  flux density of the source at $327~{\rm MHz}$ and $1.4~{\rm GHz}$, obtaining $\alpha_r=log(S_{\nu_2}/S_{\nu_1})/log({\nu_2}/{\nu_1})=-0.98$, which is in agreement with the result obtained from the spectral index map. Throughout the paper we will use $\alpha_r=-(1.0\pm0.2)$ as the average spectral index of the lobes and $\alpha_r=-(0.7\pm0.2)$ as the average spectral index of the hotspots.

The total radio luminosity of the source, obtained by integrating the synchrotron spectrum from $1~{\rm MHz}$ to $100~{\rm GHz}$ \citep[e.g., eq. 1 in][]{smolcic07},  is $L_{\rm r}=6.7 \cdot 10^{38}~{\rm W}$.

\subsubsection{Magnetic field strength in the lobes and hotspots}\label{sec:mag}
Based on the obtained radio properties of our source, we estimate the magnetic field strength in its lobes and hotspots. 
This will be further used to estimate the expected IC emission (Sec.~\ref{sec:XrayIC}).

We follow \citet{smolcic07} and apply the minimum energy condition. Briefly, the minimum energy condition corresponds almost to equipartition between the relativistic particles (protons and electrons, $E_{\rm p,e}$) and the magnetic field, $E_{B_{eq}}$ \citep[for details 
see][]{pacholczyk70}. The magnetic field strength is then given by \citep[e.g.][]{miley80}: 
\begin{eqnarray}\label{eq:mag}\nonumber
B_{\rm eq}&=&5.69\cdot10^{-5}\left(\frac{1+k}{\eta}\frac{S_r (1+z)^{3-\alpha_r}}{\nu^{\alpha_r}\theta_x\theta_y s}\right.\\
&& \left. \frac{\nu_2^{\alpha_r+1/2}-\nu_1^{\alpha_r+1/2}}{\alpha_r+\frac{1}{2}}\right)^{2/7}~{\rm G}
\end{eqnarray}
where $z$ is the redshift of the source, $S_{\rm r}~[{\rm Jy}]$ is the observed radio flux of the emitting region at frequency $\nu_{\rm r}~[{\rm GHz}]$, 
$\theta_{x,y}~[{\rm arcsec}]$ is the angular size of the emitting region, $s~[{\rm kpc}]$ is the path length through the source along the line of sight,
$\nu_{1,2}~[{\rm GHz}]$ are lower and upper frequency cutoffs, and $\alpha_r$ is the spectral index. Equation~\ref{eq:mag} assumes (i) 
a cylindrical symmetry of the lobes, (ii) equal energy densities carried by protons and electrons in the lobes ($k=1$), (iii) no relativistic 
beaming ($\delta=1$), (iv) the volume of the lobes to be completely filled with the plasma ($\eta=1$), and (v) the magnetic field is transverse to the line of sight ($\sin\Phi=1$).
Given the averaged properties of the lobes ($S_{\rm 1.4~GHz}=(29.0\pm0.5)~{\rm mJy}$, $\alpha_r=-(1.0\pm0.2)$, $\theta_x=23.8''$, $\theta_y=15.35''$, and $s=130~{\rm kpc}$) and taking 
$1~{\rm MHz}$ and $100~{\rm GHz}$ for $\nu_{1,2}$, the obtained magnetic field strength is $B_{\rm eq}=(10\pm2)~{\rm \mu G}$. The magnetic energy density is $u_B=B_{\rm eq}^2/8\pi=(4.0\pm1.5)\cdot10^{-12}~{\rm erg~cm^{-3}}$. The errors are propagated from uncertainties on $S_{1.4~{\rm GHz}}$, $\alpha_r$, and $z$.

Given the averaged properties of the hotspots ($S_{\rm 1.4~GHz}=(27.4\pm0.5)~{\rm mJy}$, $\alpha_r=-(0.7\pm0.2)$, $\theta_x=1.4''$, $\theta_y=1.5''$, and $s=12~{\rm kpc}$) and taking 
$1~{\rm MHz}$ and $100~{\rm GHz}$ for $\nu_{1,2}$, the obtained magnetic field strength is $B_{\rm eq}=(53\pm2)~{\rm \mu G}$ and the magnetic energy density is $u_B=B_{\rm eq}^2/8\pi=(113\pm8)\cdot10^{-12}~{\rm erg~cm^{-3}}$.

\subsection{Optical properties}\label{sec:opti}
\begin{figure}
\centering \includegraphics[width=.45\textwidth]{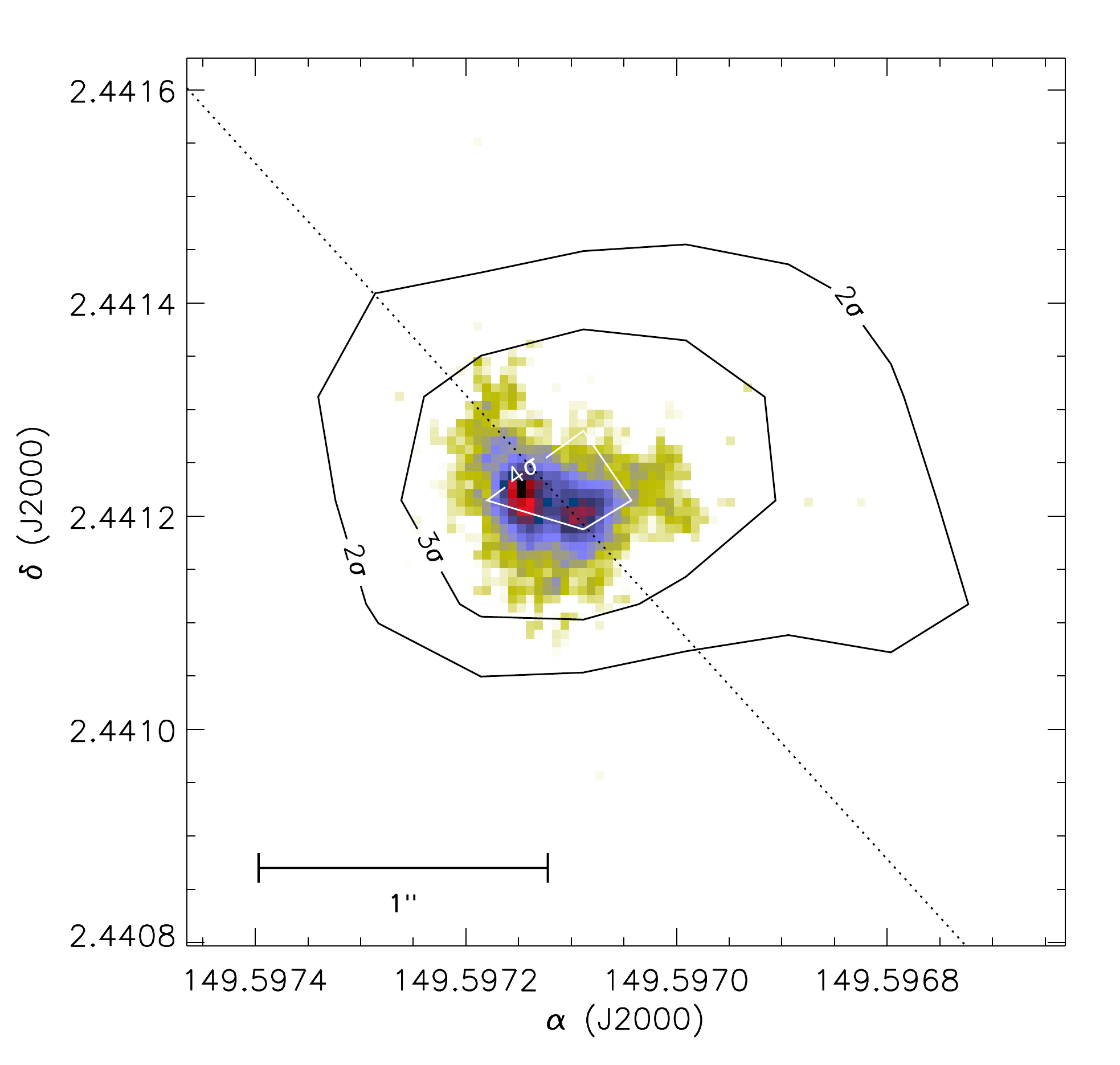}
\caption{HST-ACS {\emph i}-band image of the region around the radio core of the EC dominated X-ray candidate. The radio emission at $1.4~{\rm GHz}$
is plotted with solid contours ($\sigma=17.16~{\rm \mu Jy}/beam$). The synthesized beam of the radio image is $1.5''\times1.4''$.}
\label{fig:XlobeZoom}
\end{figure}

The radio core of the large radio galaxy is associated with an optical 
source located at $\alpha=9{\rm h}~58{\rm m}~23.31{\rm s}$,
$\delta=+02^\circ~26'~28.33''$ (see Fig.~\ref{fig:XlobeZoom}). 

In the HST-ACS image \citep{koekemoer07}, with a pixel scale 
of $0.03''$ per pixel, two central cores are clearly visible (see Fig.~\ref{fig:XlobeZoom}). 
The separation of the cores is $0.22''$, which corresponds to a physical distance 
of $\sim1.7$~kpc at the redshift of the source. The presence of two cores could be either interpreted  
as the two nuclei of merging galaxies or as two star-formation regions in a single galaxy. Alternatively,
 they are a superposition of two nuclei of two independent galaxies in the same group.
Given the average density in the group selected by Voronoi tessellation (see Sec.~\ref{sec:group};
$2.68\cdot10^5$ galaxies per $deg^2$), such a probability is, however, small, i.e., $P=3.2\cdot 10^{-3}$.

\begin{figure}
\centering \includegraphics[width=.45\textwidth]{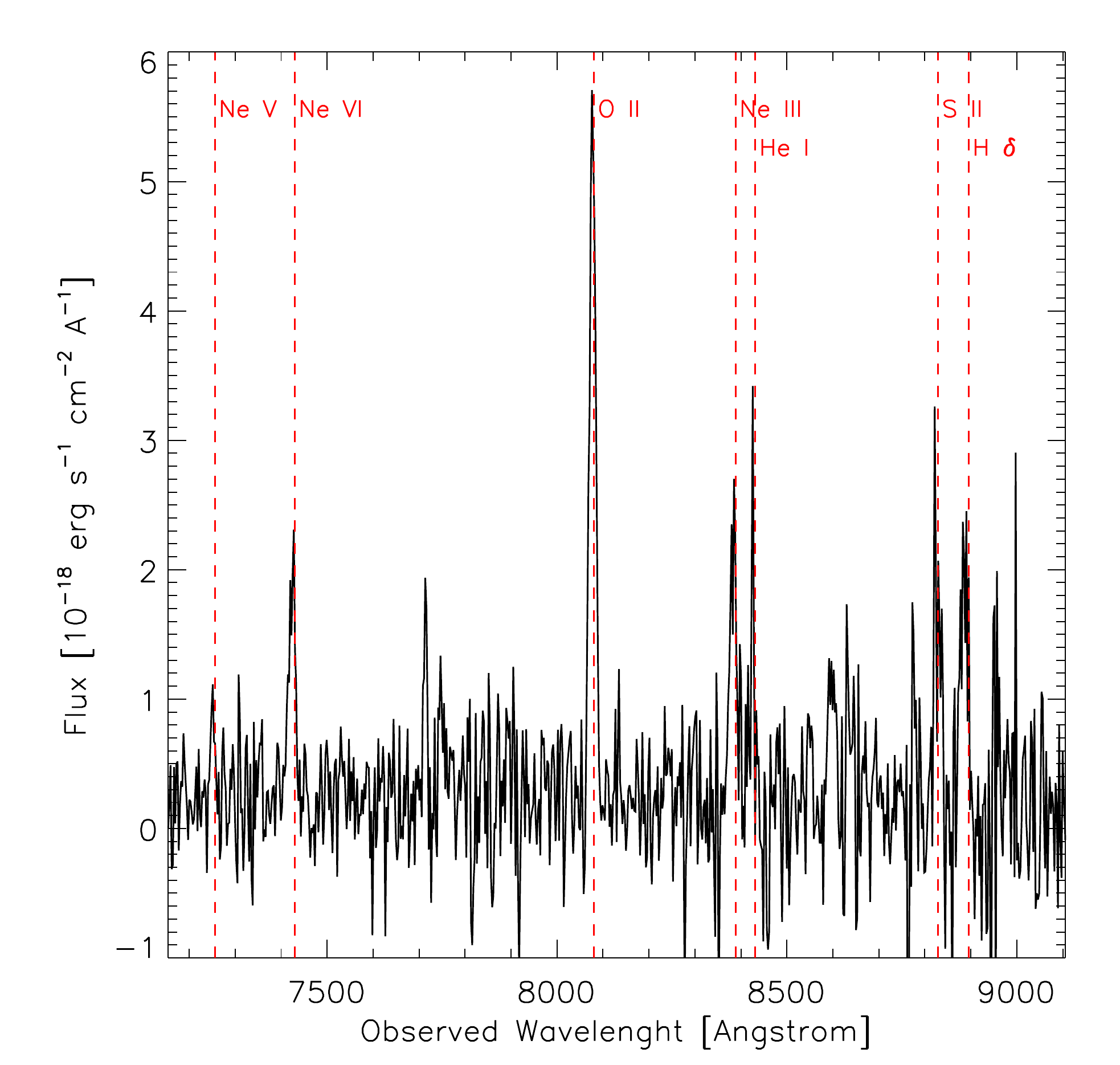}
\caption{A part of the optical spectrum of the source taken within the zCOSMOS survey. The spectroscopic
 redshift is $1.1684\pm0.0007$, and given the presence of only narrow lines ([NeV] doublet and [OII]), 
 the source is classified as a narrow line emission galaxy.}
\label{fig:spectrum}
\end{figure}

An optical spectrum (5500-9500 \AA) of the source has been taken within the zCOSMOS
survey \citep{lilly07,lilly09}. The zCOSMOS slit was oriented North-South, 
with the dispersion axis in East-West direction. The spectroscopic redshift is $1.1684\pm0.0007$, 
and given the presence of only narrow lines ([NeV] doublet at 3346 and 3426\AA\ 
and [OII] at 3727\AA, see Fig.~\ref{fig:spectrum}), the source is classified as a narrow line emission galaxy.
Given the redshift of the source, it is not possible to use a diagnostic diagram 
\citep{kewley01, lamareille04} to classify the ionizing source 
(nuclear vs. star-formation). We analyzed the optical spectrum to look for two emission line systems 
corresponding to the two optical cores in the HST-ACS image. Fitting the [NeV] emission line
 at 3426 \AA\, two peaks in the line profile are visible with a separation comparable
 to the zCOSMOS spectral resolution ($\sim$600). In this case, the spectroscopic data available 
are not sufficient to draw firm conclusions on the nature of the two optical cores. The North-South 
separation of the 2 cores is $0.1"$, so given the angular resolution of zCOSMOS ($0.20''$ per pixel), it is not possible 
to spatially resolve the two sources in the 2D spectrum either.

\begin{table*}
  \centering
  \caption{Observed X-ray flux of the EC dominated X-ray candidate compared with expected fluxes of IC emission of the hotspots and lobes
  (see Sec.~\ref{sec:XrayIC}~\&~\ref{sec:XrayIChs}), and thermal emission of the group (see Sec.~\ref{sec:Xrayther}).}
  \begin{tabular}{lccccccc}\hline\hline 
  & \multicolumn{4}{c}{\sc observables} & &\multicolumn{2}{c}{\sc derived properties}\\ \cline{2-5} \cline{7-8} 
  & $S_{\rm 1.4~GHz}$ & $\alpha_r$ & $S_{0.5-2~{\rm keV}}$ &   $\alpha_X$ & & $B_{\rm eq}$ & $S_{0.5-2~{\rm keV}}$ \\ 
  & ${\rm [mJy]}$ &  &${\rm [10^{-15}~erg~s^{-1}~cm^{-2}]}$ & & & ${\rm [\mu G]}$ &${\rm [10^{-15}~erg~s^{-1}~cm^{-2}]}$ \\ \hline  
  {\sc radio lobes} & $58.1\pm0.5$ & $-(1.0\pm0.2)$ & $2.7\pm0.7$ & $-(1.6 \pm 0.5)$ & & $10\pm2$ & $(2.5\pm0.5)_{\rm EC}$ \\
  {\sc radio hotspots} & $54.8\pm0.5$ & $-(0.7\pm0.2)$ & & & & $53\pm2$ & $(0.04\pm0.01)_{\rm SSC+EC}$ \\
  {\sc associated group} & &  & $1.4 \pm 1.2$ && & &$(0.4\pm0.1)_{\rm thermal}$ \\

 \hline \hline
  \end{tabular}\label{tab:prop}
\end{table*}

\section{Properties of the associated galaxy group} \label{sec:group}
To determine whether an optical galaxy overdensity is associated with the X-ray--emitting
lobe candidate, we apply a Voronoi tessellation analysis \citep[e.g.][]{ebeling93} to the field surrounding our radio galaxy to identify potential galaxy group members. 
The strength of the method is that no a-priori 
assumptions need to be made about cluster/group properties, making the technique sensitive to non-symmetric (e.g. elongated) structures (see e.g.\ \citealt
{smolcic07,oklopcic10}).
We first select galaxies within $\sim 20'$ (corresponding to 10 Mpc physical distance at $z=1.1$)  from the radio galaxy that have photometric redshifts within $\Delta z 
=3{\sigma_{\delta z}}(1+z_0)$ relative to  the redshift of the radio galaxy ($z_\mathrm{0}=1.168$). Following \citet{ilbert09} we take $\sigma_
{\delta z}=0.007$ and 0.034 for galaxies with $i^+<22.5$ and 26.5, respectively. 
Note that only 1 spectroscopic redshift (i.e.\ that of the radio galaxy host) is available in the redshift range and area of interest. 

The Voronoi tessellation method considers the selected galaxies as a two-dimensional 
distribution of points (called nuclei) on a plane of the sky. The plane is then 
subdivided into polygonal regions such that a surface area of each region is minimized and contains only one 
nucleus. Since the inverse of the resulting 
surface area of each region corresponds to the effective local density of the galaxy, 
one can identify overdensities and therefore clustering of galaxies. 

Once we compute the Voronoi regions, we estimate the background density via Monte Carlo simulations \citep{botzler04,smolcic07,oklopcic10}.
The same sample of galaxies is 100 times randomly redistributed over the same area. The Voronoi tessellation is then applied 
to each generated field, and the mean local density (${\bar \rho_{bg}}$) and its standard deviation ($\sigma_{bg}$) 
are calculated. All Voronoi regions with densities $\rho \geqslant {\bar \rho_{bg}}+10\sigma_{bg}$ are 
considered as overdensities.  Note that we performed the Voronoi tessellation analysis on a much larger region than the expected group/cluster area to facilitate 
the identification of overdensities.

\begin{figure}
\centering \includegraphics[width=.45\textwidth]{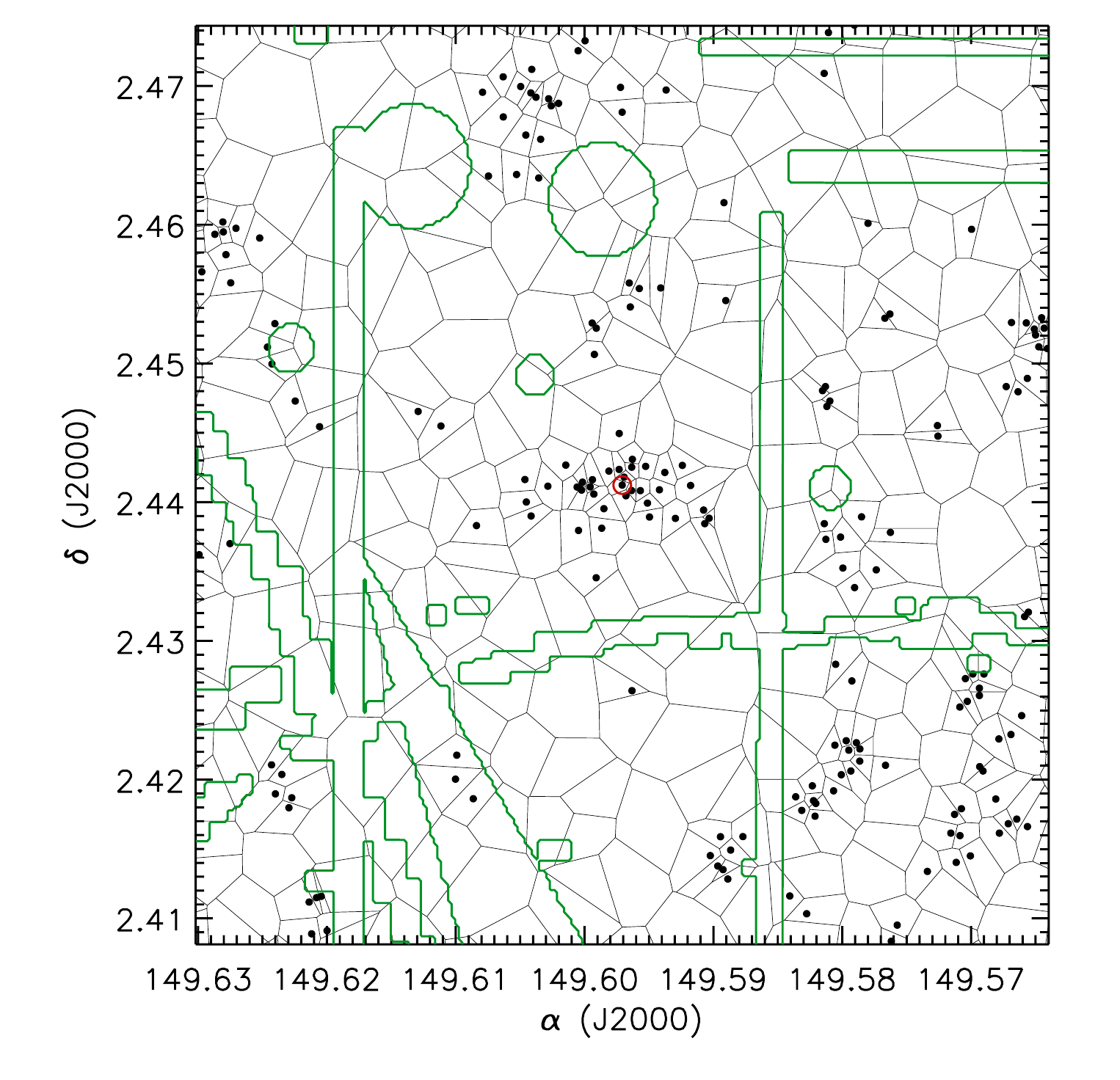}
\caption{Voronoi tessellation in the area of the EC dominated X-ray candidate. The candidate is marked with the red circle, while
the galaxies from the {\em photo-z} catalogue are shown as black dots. Green regions correspond to masked-out regions around 
saturated object in the images used for the {\em photo-z} catalogue. We have identified 37 galaxies in the immediate surrounding of the X-ray--emitting
lobe candidate.}
\label{fig:Voronoi}
\end{figure}

\begin{figure*}
\centering \includegraphics[width=.85\textwidth]{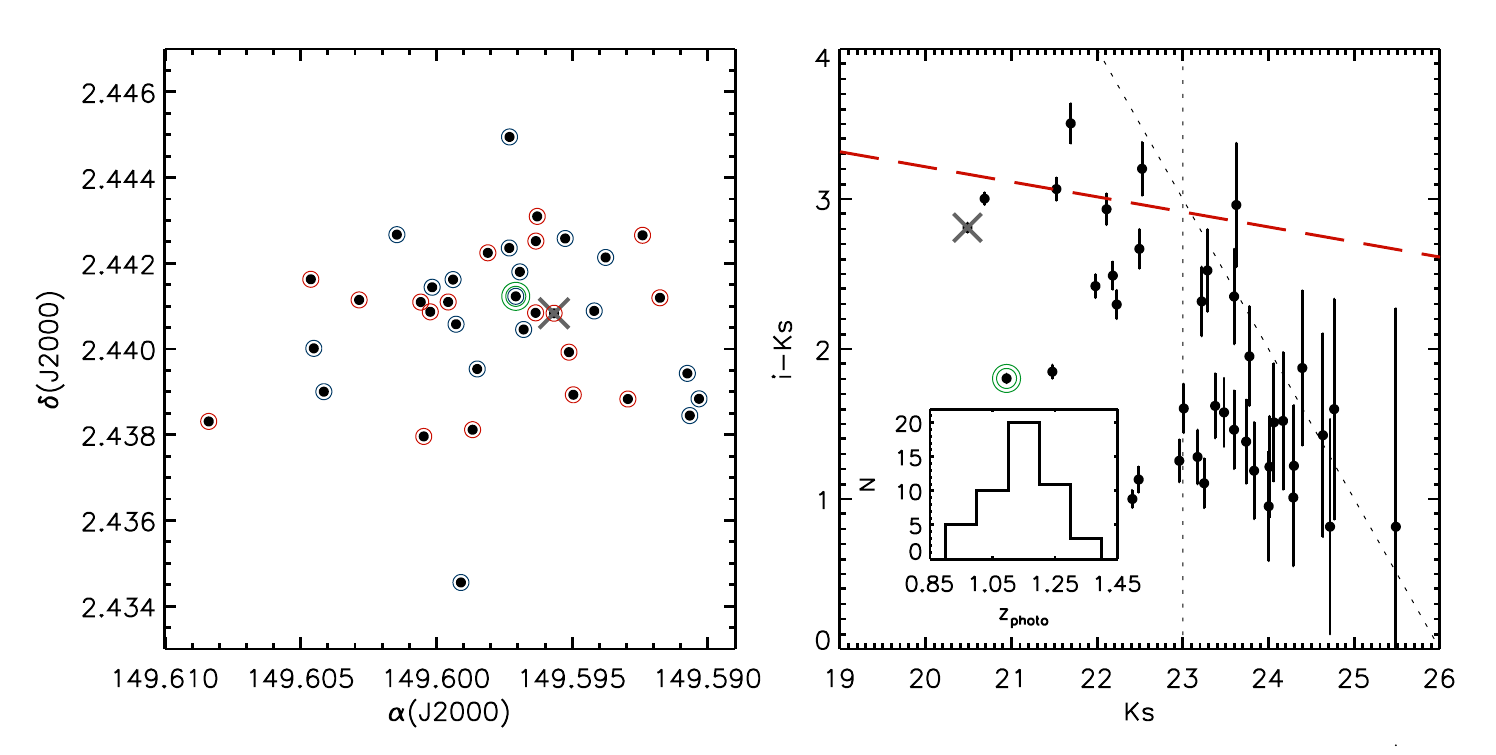}
\caption{Spatial distribution of the galaxies in the overdensity identified via a Voronoi tessellation analysis (left panel). Red ($i-K_s\geq2$) and blue ($i-
K_s<2$) galaxies are also indicated. The right panel shows the $i-K_s$ vs.\ $K_s$ color-magnitude diagram of the overdensity galaxies. A model red 
sequence expected at the group's redshift is shown by the dashed line (see text for details). The completeness limit in $K_s$ (23~mag; vertical dotted line) and $i-K_s$ color (inclined dotted line) are also shown. The brightest galaxy in the system is indicated with a cross, the optical host of the radio galaxy is circled by a double line. The photometric redshift distribution of the galaxies associated with the overdensity is shown in the inset.}
\label{fig:CMD}
\end{figure*}

The results of the Voronoi tessellation analysis, shown in Fig.~\ref{fig:Voronoi}, reveal a clear galaxy overdensity around the radio galaxy, extending over $\sim300~{\rm kpc}$. We identify 37 galaxies in the overdensity. Their redshift and spatial distribution, as well as the $i-K_s$ vs.\ $K_s$ color-magnitude diagram (CMD) are shown in Fig.~\ref{fig:CMD}. In the CMD we also show the expected red sequence at $z=1.168$. We use the \citet{bruzual03} population synthesis code with an updated treatment of thermally pulsating AGB stars.
We fitted the red sequence in nearby clusters \citep{bower92} with single burst models formed at $z=5$ and varying metallicity.  We then evolved the sequence back over time to $z=1.16$.

Both a red sequence, and a blue cloud are discernible from the data.  The brightest red sequence galaxy, which is also the most massive galaxy in the system ($M_* = 2.5\times10^{11}~\mathrm{M_\odot}$) is located close to the center of the overdensity.  The host of the radio galaxy is the third brightest galaxy, and it has a green (i.e.intermediate between red and blue) color. Such a color is consistent with that expected for powerful radio galaxies \citep{smolcic09a}, that often show signs of galaxy-galaxy mergers \citep[e.g.][see also Sec.~\ref{sec:opti}]{baum92,baldi08}. 

In order to investigate the spatial distribution of different types of galaxies in the identified overdensity we divide the galaxies into red ($i-K_s\geq2$) and blue ($i-K_s<2$) galaxies. The first can be considered to be old passive galaxies with higher stellar masses ($<M*>\approx2\times10^{10}$~M$_\odot$), while the latter are younger galaxies with likely ongoing star formation and on average lower stellar masses ($<M*>\approx5\times10^{9}$~M$_\odot$). As shown in Fig.~\ref{fig:CMD} there is no structured distribution that would be expected for relaxed groups or clusters (i.e.\ centrally concentrated massive red galaxies with less massive blue galaxies at the outskirts). The massive red galaxies are rather spread over the system, with possible higher clustering in two clumps, i) surrounding the brightest and most massive galaxy, and ii) to the East of the brightest galaxy where an agglomeration of 5 red galaxies is present. Such properties suggest a non-relaxed state of the system, i.e. an agglomeration of galaxies in the process of forming a relaxed group/cluster. 
Summing up the stellar masses of all galaxies identified in the overdensity we find a total stellar mass of $10^{12}~\mathrm{M_\odot}$. We 
conclude that the identified system is likely a dynamically young group in the process of relaxation.

\section{What causes the X-ray emission ?}\label{sec:Xrayexp}
In order to reveal the origin of the X-ray emission of the EC dominated X-ray candidate, in this section 
we calculate the expected X-ray fluxes arising from the radio lobes and hotspots, and separately from the gas of the optically identified group. We discuss the implications of the results in Sec.~\ref{sec:Xray?}.

There are two processes involving IC scattering that can result in  X-ray emission from a radio source: 
(i) synchrotron self-Compton (SSC) emission, which results from IC scattering of synchrotron 
radiation by the same relativistic electrons that produce the synchrotron radiation; 
and (ii) IC scattering of CMB photons on synchrotron emitting electrons (external Compton, EC). 
In the high-luminosity hotspots, where the electron density is high, the dominant process is usually SSC \citep{hardcastle04}. 
However, low-luminosity hotspots show X-ray emission that is much brighter than would be expected by SSC process
indicating addition component of the X-ray emission \citep{hardcastle04}. In the lobes the electron density is much lower than in the hotspots, 
so EC scattering typically dominates over  SSC \citep{croston04, croston05, kataoka05}.

\subsection{IC scattering in the radio lobes}\label{sec:XrayIC}
To estimate the EC emission coming from the lobes, we follow \citet{kataoka05}
and make the following assumptions: (i) the minimum energy condition; (ii) lobes have a cylindrical shape  
with radius $R$ and length $l$; (iii) equal energy densities carried by protons 
and electrons in the lobes ($k=1$); (iv) no relativistic beaming ($\delta=1$); and (v) the 
volume of the lobes to be completely filled with the plasma ($\eta=1$). Then, we write the ratio 
between the radio and EC X-ray luminosity, which equals to the ratio between the CMB photon energy
density and the magnetic field energy density (both calculated in the rest frame of the emitting region),
i.e., $L_{\rm EC}/L_{\rm r}\simeq u_{\rm CMB}/u_{B}$ \citep[e.g.][]{rybicki86}. The CMB photon energy density 
in the rest frame is $u_{\rm CMB}=4.1\cdot10^{-13}(1+z)^4~{\rm erg~cm^{-3}}$  \citep[e.g.][]{kataoka05} and
$u_B$ is calculated in Sec.~\ref{sec:mag}.

However, to be able to compare the predicted EC emission with the observed X-ray emission, 
we need to estimate the flux arising from the EC process in the observed 0.5--2 keV band. In the EC process,
electrons upscatter CMB photons to frequencies peaked at $\nu_{EC}$, which is, in the Thomson regime
and the frame of observer, given by \citep[e.g.][]{rybicki86, wilson09}:
\begin{equation}\label{eq:nuEC}
\nu_{EC}=\frac{4}{3}\gamma^2\nu_{\rm CMB}(1+z)^{-1},
\end{equation}
where $\nu_{\rm CMB}=1.6\cdot10^{11}(1+z)~[\rm Hz]$ is the frequency of the CMB photons, and
\begin{equation}\label{eq:gammaEC}
\gamma=\sqrt{\frac{2\pi m_e c }{e}\frac{\nu_r(1+z)}{B}}
\end{equation}
is the Lorenz factor of electrons emitting synchrotron radiation at frequency $\nu_r$ 
($m_e=9.109\cdot10^{-31}~{\rm kg}$ is the electron mass and $e=1.602\cdot10^{-19}~{\rm C}$
is the elementary charge). Then, assuming a power law with a spectral index $\alpha_X\simeq\alpha_r$ (see Sec.~\ref{sec:radio}), 
we calculate the X-ray flux, $S_{\rm X}$, at frequency $\nu_{X}$, by extrapolating the flux $S_{\rm EC}$ 
at frequency $\nu_{\rm EC}$ \citep{stawarz03}:
\begin{equation}\label{eq:fracIC}
\frac{L_{\rm EC}}{L_{\rm r}}=\frac{\nu_{\rm EC}S_{\rm EC}}{\nu_{\rm r}S_{\rm r}}=\frac{\nu_{\rm X}S_{\rm X}}{\nu_{\rm r}S_{\rm r}} \left(\frac{\nu_{\rm EC}}{\nu_{\rm X}} \right)^{1+\alpha_r}\simeq
		\frac{u_{\rm CMB}}{u_{B}},
\end{equation}
where $S_{\rm r}$ is the observed radio flux at frequency $\nu_{\rm r}$.
Finally, we integrate $S_{\rm X}$ obtained from Eq.~\ref{eq:fracIC} 
over the observed X-ray band ($0.5$ to $2~{\rm keV}$).

Given the observed/derived properties of the lobes ($\gamma=(1.0\pm0.2)\cdot10^4$,
$\nu_{EC}=(2.3\pm0.7)\cdot10^{19}~{\rm Hz}$, and see Table~\ref{tab:prop}),
the resulting integrated EC flux of the lobes in the $0.5$ to $2~{\rm keV}$ band is
$(2.5\pm0.5)\cdot10^{-15}~{\rm erg~s^{-1}~cm^{-2}}$. 
The errors are propagated from uncertainties on $S_{1.4~{\rm GHz}}$,
$\alpha_r$, $z$, and $B_{eq}$. 

\subsection{IC scattering in the radio hotspots}\label{sec:XrayIChs}
To estimate the SSC and EC emission coming from the hotspots, we follow the same procedure as
in Sec.~\ref{sec:XrayIC} and make the same assumptions (i)--(v). Note that for the hotspots we 
do not assume a cylindrical shape but a spherical shape with radius $R$. For the SSC 
emission Eq.~\ref{eq:fracIC} becomes:
\begin{equation}\label{eq:fracSSC}
\frac{L_{\rm SSC}}{L_{\rm r}}=\frac{\nu_{\rm SSC}S_{\rm SSC}}{\nu_{\rm r}S_{\rm r}}=\frac{\nu_{\rm X}S_{\rm X}}{\nu_{\rm r}S_{\rm r}} \left(\frac{\nu_{\rm SSC}}{\nu_{\rm X}} \right)^{1+\alpha_r}\simeq
		\frac{u_{\rm syn}}{u_{B}},
\end{equation}
where $\nu_{\rm SSC}=4/3\gamma^2\nu_{\rm R}$ and $u_{\rm syn}$ is the synchrotron photon energy density given in the rest frame of the emitting region by $u_{\rm syn}=(d^2_L \nu_{\rm r} S_{\rm r})/(R^2c)$.

Given the observed/derived properties of the hotspots ($\gamma=(4.5\pm0.1)\cdot10^4$,
$\nu_{SSC}=(4\pm1)\cdot10^{16}~{\rm Hz}$, $\nu_{EC}=(4\pm2)\cdot10^{18}~{\rm Hz}$, and see Table~\ref{tab:prop}),
the resulting integrated X-ray flux of the hotspots in the $0.5$ to $2~{\rm keV}$ band is 
$(8\pm2)\cdot10^{-19}~{\rm erg~s^{-1}~cm^{-2}}$ for SSC emission and $(4\pm1)\cdot10^{-17}~{\rm erg~s^{-1}~cm^{-2}}$
for EC emission. The errors are propagated from uncertainties on $S_{1.4~{\rm GHz}}$, $\alpha_r$, $z$, and $B_{eq}$.

\subsection{Thermal gas emission of the group}\label{sec:Xrayther}

In Sec.~\ref{sec:group} we have identified a galaxy overdensity around the large radio galaxy. Here we predict the X-ray flux arising from the hot gas of the group in the following way. 
We scale the total stellar mass of the group $M_*=10^{12}~\mathrm{M_\odot}$, estimated in Sec.~\ref{sec:group} via the Voronoi tessellation analysis, to the total mass $M_{500}$ (i.e., the total mass within the radius at which the density
is 500 times the critical density) using a robust correlation \citep{giodini09}\footnote{\citet{giodini09} combined stellar mass estimates in 118 galaxy groups in the COSMOS field with the weak lensing measurements of the group total mass obtained by \citet{leauthaud10}.}:

\begin{equation}
M_*=A\left( \frac{M_{500}} {5\cdot 10^{13}~{\rm h^{-1}_{72}}}\right)^{-\alpha},                                                                                                                                                                                                                                                                                                                                                                                                                                                                                                                                                                                                                                                                                                                                                        
\end{equation}
where $\log_{10} A={0.3\pm0.02}$ and $\alpha=0.81\pm0.11$. The resulting total mass of the group is $M_{500}=(2.1\pm0.3)\cdot10^{13}M_{\sun}$.

Then, we calculate the X-ray luminosity of the group based on the  luminosity-mass $L_X$--$M$ relation obtained via a weak lensing analysis in the COSMOS field \citep
{leauthaud10},
\begin{equation}
\frac{M_{200}E(z)}{10^{13.7}~{\rm M_{\sun}}}=B \left( \frac{L_{\rm X} E(z)^{-1}}{10^{42.7}~{\rm erg~s^{-1}}}\right)^{\beta},                                                                                                                                                                                                                                                                                                                                                                                                                                                                                                                                                                                                                                                                                                                                                     
\end{equation}
where 
$\log_{10} B=0.106\pm0.0053$,  $\beta=0.56\pm0.12$, $E_z=\sqrt{\Omega_m (1+z)^3+\Omega_{\Lambda}}=1.866$
for the cosmology assumed here and $z=1.168$, and $M_{500}$ is the total mass within a radius encompassing 
$\geq500$ times the critical density. Note that for the previous calculation $M_{200}$ (the mass within a radius at
200 times the critical density) was used. We can convert $M_{500}$ to $M_{200}$ assuming a NFW profile with a 
constant concentration parameter ($c=5$).  This yields an X-ray luminosity of $(1.1\pm0.3)\cdot10^{43}~{\rm erg~s^{-1}}$. Systematic
uncertainty in estimating this number is dominated by the scatter in the stellar mass - total mass relation and is a factor of 2.
The resulting X-ray flux in the $0.5$--$2~{\rm keV}$ band, due to hot gas emission of the group, 
is then $(4\pm1)\cdot10^{-16}~{\rm erg~s^{-1}~cm^{-2}}$ in the circular area centered on the radio galaxy and encompassing the radio lobes.
When calculating the flux we took into account the $L_{X}-T$ relation to derive the K-correction, following the prescription in Finoguenov et al. (2007).
This is consistent with the marginal X-ray detection of the group emission of $14\pm12\cdot10^{-16}~{\rm erg~s^{-1}~cm^{-2}}$, reported in Sec.~4.1.

\subsection{Comparison of thermal and non-thermal X-ray emission}\label{sec:Xray?}
\begin{figure}
\centering \includegraphics[width=.45\textwidth]{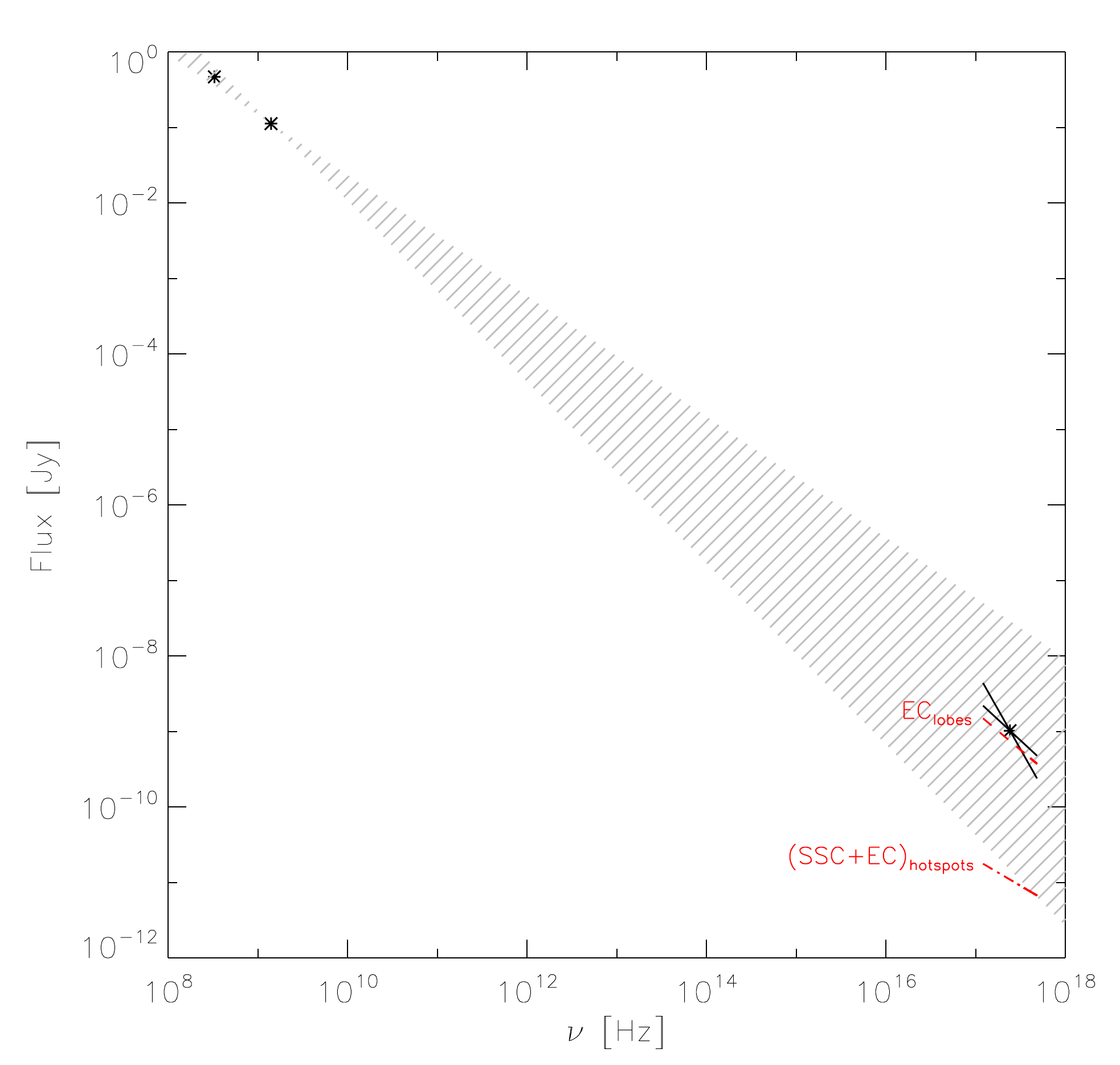}
\caption{Radio to X-ray spectral energy distribution for the EC dominated X-ray source with predicted EC emission of the lobes and SSC+EC emission of the hotspots,
showing that the measured flux is close to the predicted level for EC emission of the lobes at equipartition. The shaded area illustrates the
agreement between the X-ray and radio spectral index as expected for the EC emission.}
\label{fig:XRspectrum}
\end{figure}

The observed X-ray flux of the EC dominated X-ray candidate is compared in Table~\ref{tab:prop} to the fluxes expected from
 EC emission of the lobes and hotspots (see also Fig.~\ref{fig:XRspectrum}), as well as the thermal emission of the group. From Table~\ref{tab:prop}, one can see that 
 thermal emission of the group is almost an order of magnitude lower than what is observed. Therefore, it is very likely that 
 the observed X-ray emission of our EC dominated X-ray candidate is mostly produced by the EC process in the lobes
 with only a small contribution arising from the thermal emission of the group. The sum of the two agrees 
 within the error with the observed emission. Note that the total flux expected from SSC and EC emission of the 
 hotspots is two orders of magnitude smaller than the observed emission and thus can be ignored.
 
As presented in Sec.~\ref{sec:Xray}, the EC dominated X-ray candidate clearly shows 
extended X-ray emission associated only with the lobes of the radio galaxy, and exhibiting a
hard X-ray spectrum.  Both of these results support the idea that the X-ray emission from
 the lobes of the radio galaxy predominantly arises from non-thermal IC scattering, rather than thermal group emission.

Based on our optical analysis we have unambiguously associated a galaxy overdensity with the large radio galaxy, 
i.e. the EC dominated X-ray candidate. Our results suggest that the distribution of the group galaxies is irregular, i.e.\ inconsistent
 with that in relaxed groups or clusters. Furthermore, the optical host of the radio galaxy is found to be a 
likely merging (double nucleus) source. This resembles the case of 3C~356 at $z=1.08$ \citep{simpson02}.
 Identifying two sub-clusters associated with 3C~356, and combining their results with the studies of other 
 powerful radio galaxies, \citet{simpson02} suggested that triggering of powerful radio galaxies (at least at $z\sim1$) 
 is related to galaxy-galaxy interaction which can be coordinated by sub-cluster mergers. The unrelaxed state of the
  X-ray lobe overdensity and the double-nucleus host of the radio galaxy seem to be consistent with this scenario. 

To date 6 extended X-ray sources, where the X-ray emission is not dominated by thermal group/cluster emission,
at $z>1$ have been studied in deep X-ray surveys \citep{geach07,vardoulaki08, tu09, finoguenov10}. As the
CMB energy density is proportional to $(1+z)^4$, the number of extended X-ray sources emitting via inverse 
Compton scattering of CMB photons off the relativistic synchrotron electrons is expected to rise with redshift. 
This may affect samples of X-ray selected galaxy clusters/groups in deep X-ray surveys, which are regularly 
based on extended X-ray emission. The contribution of extended non-thermal (EC) sources to such samples
is still an open issue. This is addressed in Sec.~\ref{sec:expcounts}.

\section{Expected X-ray--counts from synchrotron lobes}\label{sec:expcounts}

\citet{celotti04} have studied the contribution of IC-related X-ray emission in the universe in comparison to that of cluster/group emission. They have shown 
that IC scattered X-ray emission may dominate that of galaxy clusters/groups above redshifts of 1 and X-ray luminosities of $10^{44}$~erg/s. In the $2\Box^\circ$ COSMOS field, searching over all ($\sim300$) extended X-ray sources with X-ray flux $S_X\geq 6\times10^{-16}$~erg/s/cm$^2$ we have found only one X-ray candidate 
where the X-ray emission may arise from IC scattering, rather than thermal ICM emission. In the following we compute how many of such sources we would 
expect in the COSMOS field, and generally in deep X-ray surveys.

First we need to determine the radio luminosity function (LF) to use for this analysis. \citet{willott01} have generated radio LFs for the two main types of radio AGN -- high radio-power (predominantly FR~IIs; high-excitation sources) and low radio-power (predominantly low-excitation FR~I and FR~IIs). The shape of the LFs, as well as their evolution is significantly different for the two (see \citealt{willott01} for details; see also \citealt{smolcic09a}). In order to estimate the number of expected extended X-ray sources on the sky, we start with the radio LF that best describes extended radio AGN, i.e. large radio galaxies. 

In deep radio surveys such as the 20~cm VLA-COSMOS, that sample well the low-power radio AGN, the fraction of large radio galaxies relative to unresolved sources (at a resolution of $1.5''$) is only $\sim10\%$ \citep{schinnerer07, smolcic08, smolcic09a}. Furthermore, the mean radio power of these large radio galaxies is $\sim9\times10^{24}$~W/Hz, i.e.\ much higher than that for unresolved sources ($\sim3\times10^{23}$~W/Hz; see e.g. \citealt{schinnerer07,smolcic09a}). On the other hand, in shallower surveys that sample predominantly high radio-power AGN, such as e.g. the Third Cambridge Survey (with the average radio luminosity of sources from the 3CRR catalog of $10^{27}$~W/Hz/sr at 151 MHz) the fraction of large radio galaxies relative to compact sources is substantially higher, i.e.\ $\sim95\%$. Thus, we adopt the luminosity function of powerful radio AGN from \citet{willott01} to compute the number of extended X-ray sources on the sky.

\begin{figure}
\centering \includegraphics[width=.45\textwidth]{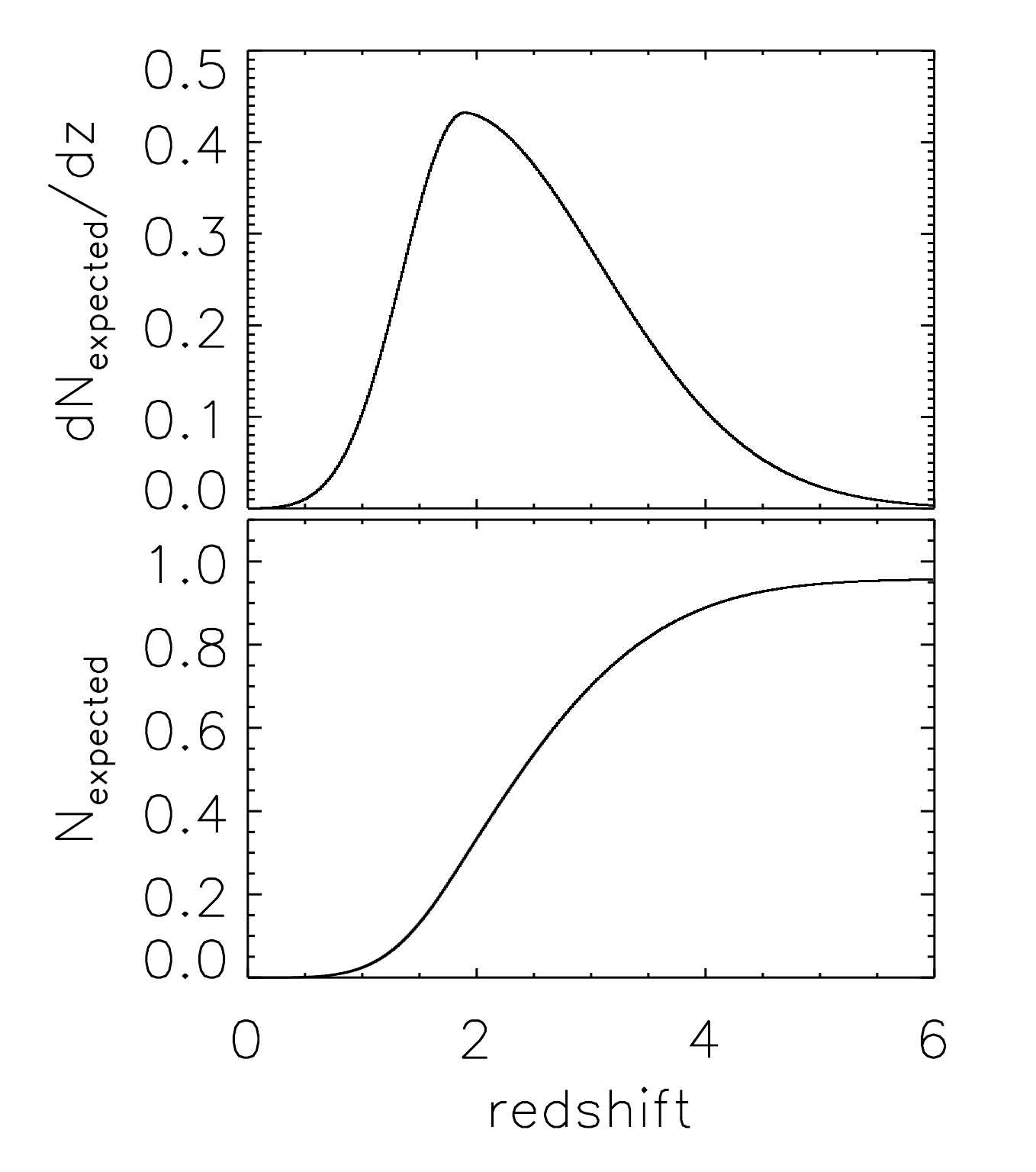}
\caption{The top (bottom) panel shows the expected differential (cumulative) number of X-ray-emitting
lobes in the COSMOS $2\Box^\circ$ field, drawn from the evolving luminosity function of powerful radio AGN (\citealt{willott01},
covering the range of  $10^{25}$ to $10^{30}$ ${\rm W/Hz}$; see text for details).}
\label{fig:counts}
\end{figure}

We first compute the expected number of extended X-ray sources due to IC scattering of CMB photons (EC) in a $2\Box^\circ$ field (such as COSMOS) by evolving the LF for powerful radio AGN covering the range of  $10^{25}$ to $10^{30}$ ${\rm W/Hz}$ (\citealt{willott01}; see also \citealt{smolcic09a}). For each redshift we convert the radio luminosities to radio flux assuming a typical spectral index of $-0.75$. Assuming that the X-ray 
emission is caused by the EC process, we then compute the expected X-ray flux arising from EC-scattering as described in Sec.~\ref{sec:XrayIC}. Given our COSMOS detection thresholds in radio and X-rays, we then calculate
the expected number of sources that could be detected above a radio flux of $S_r \ge 50~\mu$Jy (VLA-COSMOS detection limit) and an X-ray flux of $S_X 
\ge 6\times10^{-16}$~erg/s/cm$^2$ (COSMOS X-ray group detection limit) in  a $2\Box^\circ$ field. The differential and integrated number of expected extended EC X-ray 
sources is shown in Fig.~\ref{fig:counts}. In a field like COSMOS with deep radio and X-ray observations, only one X-ray source where the emission 
arises from IC-scattered electrons of the CMB is expected. This is consistent with our systematic search for non-thermal extended X-ray sources which yielded 
only one EC dominated source (discussed in detail here).
  
\begin{figure}
\centering \includegraphics[width=.45\textwidth]{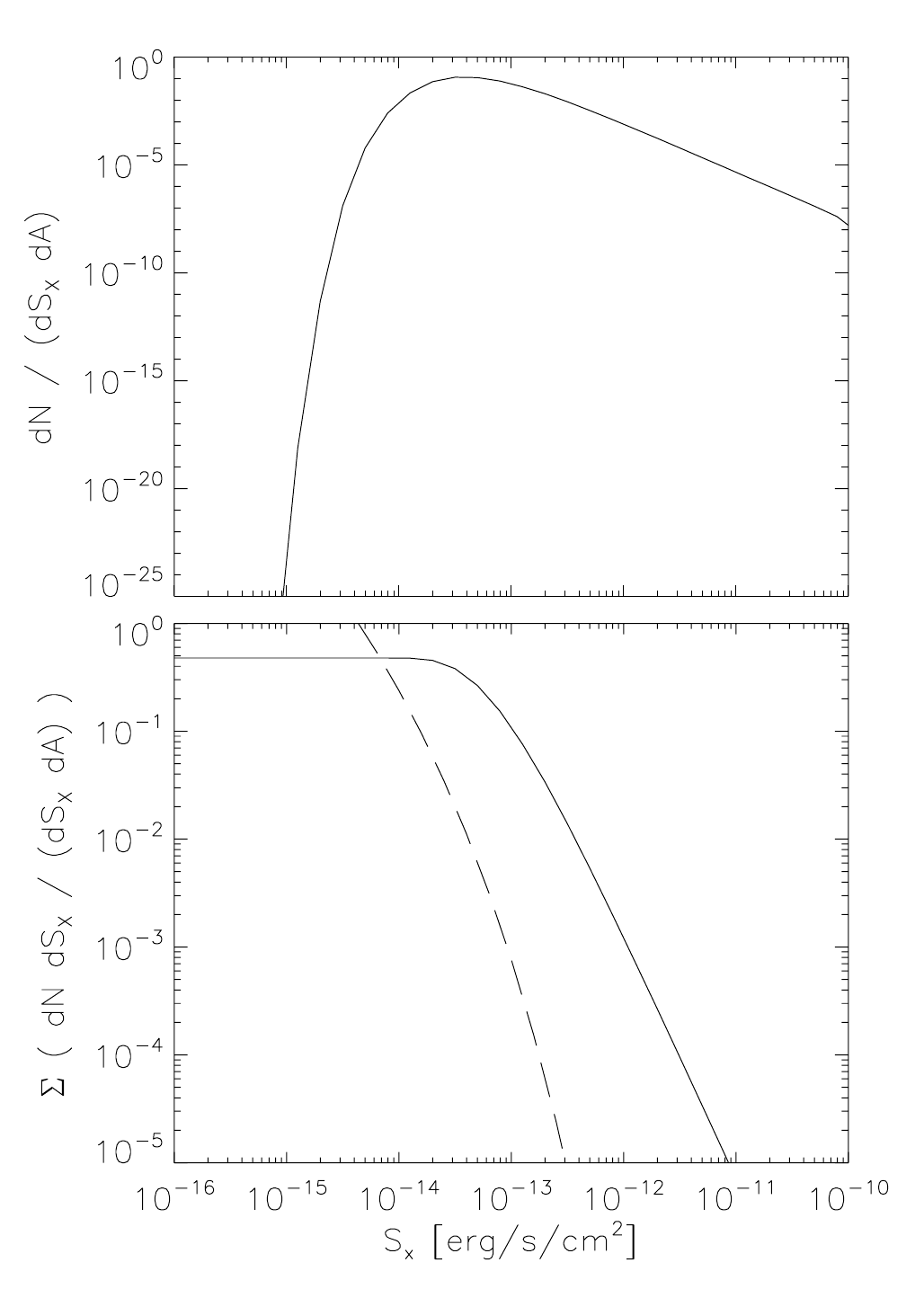}
\caption{Differential (top panel) and cumulative (bottom panel) number counts (per $\Box^\circ$) as a function of X-ray flux expected for extended inverse Compton X-ray sources (solid lines) and for clusters (dashed line). The radio counts were obtained by evolving the radio luminosity function for powerful (FR~II) radio galaxies (\citealt{willott01}, covering the range of  $10^{25}$ to $10^{30}$ ${\rm W/Hz}$; see text for 
details), while the cluster counts, in the redshift range of $1<z<3$, were obtained  following \citet{finoguenov10}.}
\label{fig:logNlogS}
\end{figure}

Extending this line of reasoning further we can predict the IC X-ray counts as a function of X-ray flux. Again, evolving the radio luminosity function for powerful radio 
galaxies ($10^{25}-10^{30}$~W/Hz; \citealt{willott01}) in a redshift range $0<z<6$, we compute the total expected number of EC (extended) X-ray sources per $\Box^\circ$ of sky. However, unlike in the calculation above, no radio flux limit is imposed here. The differential and cumulative counts are shown in Fig.~\ref{fig:logNlogS}.  The peak in the differential source counts (that corresponds to the flattening in the cumulative counts) at $L_X\sim5\times10^{-14}$~erg/s/cm$^2$ is due to the shape of the radio luminosity function for powerful radio galaxies that has a strong peak at $L_\mathrm{20cm}\sim10^{27}$~W/Hz (see \citealt{smolcic09a}). Thus the major contribution to the X-ray counts arises from such galaxies at redshifts 1--3. On the other hand, the number density of $z>1$ extended X-ray sources in the COSMOS field is $\sim20$ per $\Box^\circ$. Thus the expected $\sim0.5$ non-thermal X-ray extended source per $\Box^\circ$ (under the assumption that the underlying radio LF is correct) are not likely to dominate samples of extended (thermal) X-ray sources, i.e. clusters/groups in deep X-ray surveys. To investigate this further, following \citet{finoguenov10}, we have also computed the source counts for clusters in the redshift range of  $1<z<3$ (dashed line in Fig.~\ref{fig:logNlogS}). 
As one can see from Fig.~\ref{fig:logNlogS} the number of high-z clusters becomes comparable to that of high-z radio galaxies at fluxes exceeding $7\cdot10^{-15}~{\rm erg~s^{-1}~cm^2}$. 
Thus, EC dominated X-ray sources may produce a serious contamination for the upcoming eROSITA survey \citep{predehl10}.

\section{Summary and conclusions}\label{sec:con}
In the COSMOS field we have carried out a systematic search for non-thermal extended X-ray sources, i.e.\ sources in which the X-ray emission is mostly arising due to IC scattering of CMB photons off electrons in
the radio lobes (EC proccess), rather than thermal emission from the hot gas in the galaxy cluster/group. Based on a concurrence 
of morphological structures in the radio and X-ray images, we have found only one candidate.

The radio counterpart of our candidate is a large powerful radio galaxy ($L_{\rm r}=6.7\cdot10^{38}~{\rm W}$) hosted by a double-nuclei galaxy at $z=1.1684$.  The observed X-ray emission ($S_{\rm 0.5-2~keV}=(3.6\pm1.2)\cdot10^{-15}~{\rm erg~s^{-1}~cm^{-2}}$) is extremely elongated in the direction of the radio lobes (see Fig.~\ref{fig:Xlobe}), suggesting a non-group origin. 
In order to identify the origin of the X-ray emission in our source, we have performed a detailed analysis of the expected X-ray emission arising from inverse Compton emission in the radio lobes, the hotspots, as well as that expected from the group environment of the radio galaxy.  We find that External Inverse Compton emission of the lobes is the dominant process that generates the observed X-ray emission of our candidate (see Fig.~\ref{fig:XRspectrum}), with a minor contribution from the intragroup gas of the galaxy group associated with the radio galaxy.

Making use of the radio luminosity function for powerful radio galaxies, we have estimated the expected number of extended non-thermal X-ray sources (due to Inverse Compton scattering of CMB photons off the synchrotron electrons) on the sky as a function of X-ray flux. In a $2\Box^\circ$ field, such as COSMOS, we expect to find only one such source, consistent with our results. Furthermore, our analysis shows that such sources (in a redshift range $0<z<6$ and with radio luminosity between $10^{25}$ and $10^{30}$ ${\rm W/Hz}$) are not expected to be a significant contaminant of deep X-ray selected cluster/group catalogs, but they dominate the $z>1$ cluster counts at the bright end ($S_{\rm X}>7\cdot10^{-15}~{\rm erg~s^{-1}~cm^2}$).

\section*{acknowledgement}
We acknowledge the anonymous referees for their constructive comments. This research is in part funded by the European Union's Seventh
Frame-work program under grant agreement 229517 and contract PRIN-INAF 2007; by World Premier International Research
Center Initiative (WPI Initiative), MEXT, Japan; and  by KAKENHI No. 23740144. FC acknowledges the Blancheflor Boncompagni
Ludovisi foundation and the Smithsonian Scholarly Studies. 
\bibliographystyle{mn2e}
\bibliography{reflistCOSMOS}

\appendix

\bsp

\label{lastpage}

\end{document}